\documentclass[lettersize,journal]{IEEEtran}
\usepackage{amsmath,amsfonts}
\usepackage{array}
\usepackage{textcomp}
\usepackage{stfloats}
\usepackage{url}
\usepackage{verbatim}
\usepackage{cite}
\usepackage{xcolor}
\usepackage{amsmath}
\usepackage{amsfonts,amssymb,amsthm, bbm,braket}
\usepackage{graphicx}
\usepackage[ruled, linesnumbered, vlined]{algorithm2e}
\usepackage{caption}
\usepackage{subcaption}

\usepackage{amssymb}
\usepackage{tkz-euclide}

\newcommand\submittedtext{%
  \footnotesize This work has been submitted to the IEEE for possible publication. Copyright may be transferred without notice, after which this version may no longer be accessible.}

\newcommand\submittednotice{%
\begin{tikzpicture}[remember picture,overlay]
\node[anchor=south,yshift=10pt] at (current page.south) {\fbox{\parbox{\dimexpr0.65\textwidth-\fboxsep-\fboxrule\relax}{\submittedtext}}};
\end{tikzpicture}%
}

\begin{document}

\title{A Deep Unfolding-Based Scalarization Approach for Power Control in D2D Networks}

\author{\IEEEauthorblockN{
        Jan Christian Hauffen\IEEEauthorrefmark{1}, 
        Peter Jung\IEEEauthorrefmark{2}\IEEEauthorrefmark{1}, 
        Giuseppe Caire\IEEEauthorrefmark{1}}
        
        \IEEEauthorblockA{\IEEEauthorrefmark{1}Communications and Information Theory, Technical University of Berlin, Germany\\
        \IEEEauthorrefmark{2}German Aerospace Center (DLR), Berlin, Germany\\
        Emails: \{j.hauffen,peter.jung, caire\}@tu-berlin.de}}



\maketitle

\submittednotice
\begin{abstract}
Optimizing network utility in device-to-device networks is typically formulated as a non-convex optimization problem. This paper addresses the scenario where the optimization variables are from a bounded but continuous set, allowing each device to perform power control. The power at each link is optimized to maximize a desired network utility. Specifically, we consider the weighted-sum-rate. The state of the art benchmark for this problem is fractional programming with quadratic transform, known as FPLinQ. We propose a scalarization approach to transform the weighted-sum-rate, developing an iterative algorithm that depends on step sizes, a reference, and a direction vector. By employing the deep unfolding approach, we optimize these parameters by presenting the iterative algorithm as a finite sequence of steps, enabling it to be trained as a deep neural network. Numerical experiments demonstrate that the unfolded algorithm performs comparably to the benchmark in most cases while exhibiting lower complexity. Furthermore, the unfolded algorithm shows strong generalizability in terms of varying the number of users, the signal-to-noise ratio and arbitrary weights. The weighted-sum-rate maximizer can be integrated into a low-complexity fairness scheduler, updating priority weights via virtual queues and Lyapunov Drift Plus Penalty. This is demonstrated through experiments using proportional and max-min fairness.
\end{abstract}


\section{Introduction}
In this article we present a deep unfolding approach to solve the problem of maximizing the weighted-sum-rate in a dense device-to-device (D2D) network with full frequency reuse for the case where each transmitter is able to adjust its own power level. This is generally a non-convex optimization problem that amount to a form of power control with the goal of maximizing the weighted-sum-rate objective function. The objective function weights are used to enforce some form of priority or fairness for each link and can thus control the D2D network. It is known that standard fairness in terms of the long-term average user throughput, such as proportional fairness and max-min fairness, or more general $\alpha$-fairness \cite{mo2000fair}, can be obtained by maximizing at each scheduling slot the weighted sum of (instantaneous) rates, while recomputing iteratively the weights according to the virtual queue scheme of \cite{neely2010stochastic}. Power control is generally considered to achieve better sum rate compared to the simple on/off method of link activation, known as link scheduling.

Link scheduling can be approximately solved by greedy algorithms or computationally complex algorithms. In \cite{wu_flashlinq_nodate} a thresholding based approach, called FlashLinQ, is introduced. FlashLinQ decides in each time step if a link should be active or not, by checking if this link does not cause too much interference to already scheduled ones and if this link does not "suffer" too much interference from already scheduled ones. Another thresholding-based scheme is introduced in \cite{naderializadeh_itlinq_2014} by Naderializadeh et al. and is based on the information theoretic optimality condition of treating interference as noise (TIN) \cite{geng_optimality_2015}. In \cite{geng_optimality_2015} Geng et al. proved that TIN achieves the full capacity region of a $K$-user Gaussian interference network up to a constant gap if the channel gain for each link is not less than the sum of the maximum caused interference and the maximum received interference (in $\log$ scale). ITLinQ \cite{naderializadeh_itlinq_2014} then schedules at each discrete time step an \textit{information theoretic independent set}, i.e. a subset of all links, which satisfy the TIN optimality condition. In \cite{yi_optimality_2015} an improved version of ITLinQ, called ITLinQ+, is introduced on relaxed TIN conditions. A two stage approach for power control is also investigated, where at first ITLinQ+ selects a subset of active links, on which then power control in the high signal-to-interference-plus-noise regime is performed via geometric programming, \cite{chiang2007power}. In \cite{shen_fplinq_2017}, FPLinQ is introduced by reformulating the original non-convex problem into a \textit{sum-of-ratios} form, followed by the application of Lagrangian and quadratic transformations to derive a closed-form iterative algorithm. This iterative algorithm outperforms ITLinQ, ITLinQ+ and FlashLinQ in terms of weighted-sum-rate and number of simultaneously active links and is also developed for power control \cite{shen_fractional_2018, shen_fractional_2018-1}.

As these schemes are too complex or are sub-optimal, many machine learning approaches have emerged in recent years. Cui et al. propose in \cite{cui_spatial_2019} a machine learning approach based on the geographic location of links, since the CSI can be interpreted as the output of a stochastic function dependent on the geographic location information. This approach could achieve over $98\%$ of  the average sum rate, w.r.t FPLinQ, see \cite{cui_spatial_2019}. However, it could be observed that the spatial deep learning method favours links with a short distance and the trained kernel has a radial structure, which derives from the underlying assumption of radial symmetric path-loss. 

Modeling the D2D wireless network via a connectivity/conflict graph, approaches like \cite{lee_graph_2020, zhao2022link} and \cite{shelim_geometric_2022} address the non-Euclidean geometric structure of the network by applying graph or geometric machine learning. In \cite{lee_graph_2020} Lee et al. present a machine learning approach based on the feature vector of a given conflict graph to solve the weighted-sum-rate maximization by link selection, i.e., on-off power control. The proposed network first learns a low dimensional feature vector of the graph and uses this to learn a classifier to decide if a link should be active or not. They investigated supervised and unsupervised learning. In \cite{zhao2022link} a tree search guided by a graph neural network is proposed. 

It is important to note that while pathloss models with radially symmetric statistics are commonly used in wireless communications, this does not correspond to the reality of wireless networks, where devices are placed in a specific propagation environment. Various elements such as blockages of buildings, street canyons, directional antennas, and other propagation effects cause the actual pathloss, which defines the (large-scale) channel coefficient strength between any two devices, to be significantly non-radially symmetric \cite{levie_radiounet_2021}.

In \cite{shelim_geometric_2022} Shelim et al. propose a machine learning approach for link scheduling by measuring interference strength through regularized Laplacian matrices using the Riemannian metric. Then a geometric support vector machine is trained in a supervised way to classify link scheduling. In \cite{sun_learning_2018}, a deep neural network is trained in a supervised way to learn a pre-computed solution. This approach comes with a few disadvantages as working only well for a low number of links and requires a large training set. Lee et al. present in \cite{lee2018deep} an unsupervised deep neural network to for the power control problem in underlaid D2D communication. 

Since optimization algorithms (e.g., FPLinQ) are iterative, one can unfold the algorithm over a finite number of iterations and treat these stages as the layers of a deep neural network. Then, the parameters of the unfolded algorithm can be trained in a data-driven way, by stochastic gradient training. This approach is referred to as deep unfolding \cite{chen2018convlista}. Deep unfolding also attained recent attention for D2D communications. In \cite{li_graph-based_2022} an algorithm unfolding based network is proposed. This algorithm aims to solve the weighted sum energy efficiency (WSEE) by unfolding a line search solving the successive convex/concave approximation of the WSEE problem. 

The majority of machine learning methods have in common that they can only solve the weighted sum problem for weights equal to one. In \cite{cui_spatial_2019} a heuristic approach is presented to achieve proportional fairness, by thresholding weights to zero or one. In \cite{lee2018deep} also a heuristic for weights is used.  The networks proposed in \cite{shen2019graph} and \cite{sun_learning_2018} are able to learn for different weights, but in a supervised way, which requires the generation of a large training set where the network geometry and rate weight instances are labeled with the solution provided by FPLinQ. As such, these approaches can only learn the FPLinQ solution and therefore are generally outperformed by FPLinQ. 

In this paper, we begin by reformulating the weighted-sum-rate problem as the scalarization of a multi-objective optimization problem to locate a solution on the Pareto-Boundary. Next, we derive an iterative algorithm leveraging the Lagrangian method to approach the optimal point. Subsequently, we apply the deep unfolding technique to this iterative algorithm and introduce a data-driven, unsupervised method for learning the algorithm's parameters. The trained algorithm, consequently, can be efficiently executed as a deep neural network. The resulting model is very light in terms of trainable parameters, scalable to different number of users, and can handle arbitrary weights and general D2D link gains, making no assumptions on radial symmetry on the propagation pathloss. Additionally, by unfolding the iterative algorithm, we match or exceed the performance of FPLinQ while reducing computational complexity.

The paper is organized as follows. We first present the mathematical model for a $K$-user D2D network and the corresponding power control problem. We then introduce the necessary background on vector optimization and the Pascoletti-Serafini scalarization method, \cite{pascoletti1984scalarizing}, employed in this work, along with an overview of deep unfolding. Finally, we demonstrate the effectiveness of the proposed unfolded algorithm through numerical experiments. 

\noindent\textbf{Notation:} We denote matrices respectively vectors with bold capital or bold small letters, i.e. $\mathbf{A, a} $. $\| \mathbf{a} \|_2$ denotes the $\ell_2$-norm of some vector $\mathbf{a}$ and $|x|$ the absolute value of $x\in\mathbb{C}$. $\mathbf{I}_K$ for $K\in\mathbb{N}$ is the $K\times K$ identity matrix and $\mathbf{1}\in\mathbb{R}^K$ is an all ones $K$-dimensional vector, analogously $\mathbf{0}\in\mathbb{R}^K$ is a vector of all zeros. $\mathbb{R}_+^K$ is the positive orthant in $\mathbb{R}^K$ and includes all $\mathbf{x}\in\mathbb{R}^K$ with $x_i \geq 0,\, i=1,\dots, K$, $\text{int}(\mathcal{K})$ is the interior of a set $\mathcal{K}$. $\mathbb{E}[x]$ is the expected value of a random variable $x$.

\begin{figure}
    \centering
    \begin{tikzpicture}[scale=1.3, every node/.style={scale=0.8}]


\coordinate[label=below:$\text{Tx}_1$] (T_1) at (1,0);
\coordinate[label=left:$\text{Rx}_1$] (R_1) at (0,1);
\coordinate[label=above:$\text{Tx}_2$] (T_2) at (1,3);
\coordinate[label=left:$\text{Rx}_2$] (R_2) at (0,2);
\coordinate[label=above:$\text{Tx}_3$] (T_3) at (3,3);
\coordinate[label=above:$\text{Rx}_3$] (R_3) at (2,3);
\coordinate[label=below:$\text{Tx}_4$] (T_4) at (3,0);
\coordinate[label=right:$\text{Rx}_4$] (R_4) at (4,1);

\fill (T_1) circle (1pt);
\fill (T_2) circle (1pt);
\fill (T_3) circle (1pt);
\fill (T_4) circle (1pt);
\fill (R_1) circle (1pt);
\fill (R_2) circle (1pt);
\fill (R_3) circle (1pt);
\fill (R_4) circle (1pt);

\draw[-stealth] (T_1) -- (R_1) node[pos=0.5, label=left:$h_{11}$] {};
\draw[-stealth, dashed] (T_1) -- (R_2);
\draw[-stealth, dashed] (T_1) -- (R_3) node[pos=0.5, label=right:$h_{31}$] {};
\draw[-stealth, dashed] (T_1) -- (R_4);

\draw[-stealth] (T_2) -- (R_2) node[pos=0.5, label=above:$h_{22}$] {};
\draw[-stealth, dashed] (T_2) -- (R_1);
\draw[-stealth, dashed] (T_2) -- (R_3);
\draw[-stealth, dashed] (T_2) -- (R_4) node[pos=0.7, label=above:$h_{42}$] {};

\draw[-stealth] (T_3) -- (R_3) node[pos=0.5, label=above:$h_{33}$] {};
\draw[-stealth, dashed] (T_3) -- (R_1);
\draw[-stealth, dashed] (T_3) -- (R_2);
\draw[-stealth, dashed] (T_3) -- (R_4) node[pos=0.5, label=above:$h_{43}$] {};

\draw[-stealth] (T_4) -- (R_4) node[pos=0.5, label=right:$h_{44}$] {};
\draw[-stealth, dashed] (T_4) -- (R_1);
\draw[-stealth, dashed] (T_4) -- (R_2);
\draw[-stealth, dashed] (T_4) -- (R_3);

\coordinate[label=right: interference link] (d) at (3, 3.9);
\coordinate[label=right: communication link] (_) at (3, 3.7);
\draw[dashed] (2, 3.9) -- (3, 3.9);
\draw[] (2, 3.7) -- (3, 3.7);
\end{tikzpicture}
    \caption{Example of a D2D network with $K = 4$ Links.}
    \label{fig:ExampleNetwork}
\end{figure}
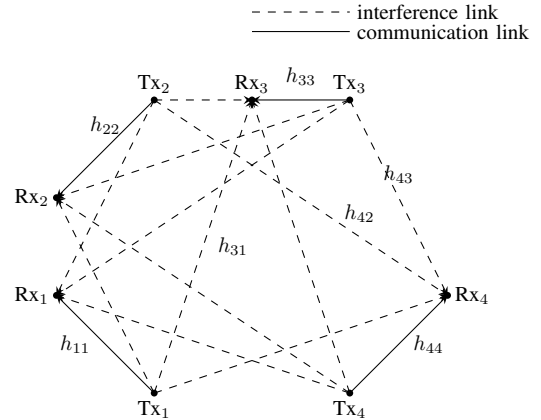

\section{Mathematical Model and Background}
In this section we state the weighted-sum-rate problem for a $K$-user D2D-network and provide the background for vector optimization used in this work, in particular, we present the scalarization approach at the basis of our proposed scheme. 
\subsection{Mathematical Model and Problem Formulation}
We consider a $K$-users D2D wireless network defined by $K$ transmitter-receiver pairs sharing the same bandwidth and geographic area. Each Tx-Rx pair is referred to as a link. The signal at the receiver of link i is given by 
\begin{align}
    Y_i(t) = \sum_{j=1}^K h_{ij}\Tilde{X}_j(t)+Z_i(t)\, , i=1,\dots,K \label{eq:K_user_channel}
\end{align} 
with power constraint $\mathbb{E}[|\Tilde{X}_i(t)|^2]\leq P_i $, where $\Tilde{X}_j(t)$ is the transmitted symbol of transmitter $j$, $Y_i(t)$ the received signal at receiver $i$ and $Z_i(t)\thicksim\mathcal{CN}\left(0,\sigma^2\right)$ for each discrete time point $t$. Here $h_{ij}\in\mathbb{C}$ is the channel coefficient from transmitter $j$ to receiver $i$, see Figure \ref{fig:ExampleNetwork}. Notice that the model in \eqref{eq:K_user_channel} corresponds to a $K$-user Gaussian interference channel \cite{el2011network}. However, for practical reasons in the great majority of works on D2D networks it is assumed that interference is treated as noise, e.g. \cite{shen_fplinq_2017, shen_fractional_2018, naderializadeh_itlinq_2014, wu_flashlinq_nodate,geng_optimality_2015, yi_optimality_2015, cui_spatial_2019, lee_graph_2020, shelim_geometric_2022, lee2018deep}. 
In this paper, we also follow this practically relevant assumption. In this case, an important quantity is the signal-to-interference-plus-noise ratio, given as
\begin{align}
	\text{SINR}_i = \frac{|h_{ii}|^2 P_i}{\sigma^2+\sum_{i\neq j}|h_{ij}|^2P_j}.
\end{align}
By treating interference as noise, each link $i$ can achieve the rate, \cite{geng_optimality_2015}:
\begin{align}
	R_i = \log\left(1+\text{SINR}_i\right).
\end{align}
The goal is now to find the optimal power level for each link, which maximizes the weighted-sum-rate $\sum_{i=1}^K w_iR_i$. We define the power control optimization variables as the vector $\mathbf{x}\in\Omega = [0,1]^K$ and a matrix $\mathbf{G}$ containing all the system parameters, $\mathbf{G}_{ij} = |h_{ij}|^2 P_j $ Then, the $i$-th user rate can be written as
\begin{align}
R_i(\mathbf{x}, \mathbf{G}) &= \log\left(1+\frac{\mathbf{G}_{ii}x_i}{\sigma^2+\sum_{i\neq j}\mathbf{G}_{ij}x_j}\right).
\end{align}
The resulting weighted-sum-rate maximization (WSRM) problem is expressed as: 
\begin{align}
\max_{\mathbf{x}\in \Omega} \sum_{i=1}^K w_iR_i(\mathbf{x}, \mathbf{G})\label{eq:weightedSum}.
\end{align}
Since by setting the variables ${\bf x} \in \Omega$ the transmit power of each transmitter $i$ is tuned from its maximum power $P_i$ to (possibly) $0$, we refer to \eqref{eq:weightedSum} as the WSRM power control problem. It is clear, that the weighted-sum-rate achieved by power control will perform better than the one obtained by link selection. Link selection reduces to an extreme form of on-off power control, i.e. $\bf x$ is a binary vector thus $\Omega = \lbrace 0,1\rbrace^K$, and has been also widely investigated in the literature for solving \eqref{eq:weightedSum}, as previously reviewed in Section I.


\subsection{Multiobjective / Vector Optimization}
In the following, we review some notations and results from \cite{eichfelder_adaptive_2008}. Typically, a vector optimization problem w.r.t. to a given convex cone $\mathcal{K}$ is stated as the following. Let 
\begin{align*}
    \Omega:= \lbrace \mathbf{x}\in\mathcal{S}\,\vert\, \mathbf{g}(\mathbf{x})\in\mathcal{C}\text{, } \mathbf{h}(\mathbf{x})={\bf 0}\rbrace
\end{align*}
be a set of constraints, where it is assumed that the given functions $\mathbf{f}: \mathbb{R}^n\to \mathbb{R}^m$, $\mathbf{g}: \mathbb{R}^n\to \mathbb{R}^p$, $\mathbf{h}: \mathbb{R}^n\to \mathbb{R}^q$ are smooth. Moreover let $\mathcal{S}\subset \mathbb{R}^n$ be a closed convex set, and $\mathcal{C}\subset \mathbb{R}^p$ a closed convex cone. We consider the following problem formulation
\begin{subequations} \label{MOP} 
\begin{align}
    \min \,& \mathbf{f}(\mathbf{x})\, \text{ w.r.t. } \,\mathcal{K}\\
    \text{subject to } & \mathbf{x}\in \Omega .
\end{align} 
\end{subequations}
The optimal solutions to \eqref{MOP} are defined through the convex cone $\mathcal{K}$. Any convex cone $\mathcal{K}$ defines a partial ordering $\leq_\mathcal{K}$ in $\mathbb{R}^m$, that is 
\begin{align*}
    {\bf x} \leq_\mathcal{K}{\bf y}\, \Leftrightarrow\, {\bf y} - {\bf x} \in\mathcal{K}.
\end{align*}
and thus decides how the optimal points of \eqref{MOP} are defined.
An optimal point $\hat{\mathbf{x}}$ of \eqref{MOP} w.r.t. $\mathcal{K}$ is referred to as a $\mathcal{K}$-optimal point. A $\mathcal{K}$-optimal point $\hat{\bf x}$ in vector optimization is a point that is non-dominated by any other point in the given set according to the partial ordering induced by the cone $\mathcal{K}$, i.e.,
\begin{align}
	\hat{\bf x}:\,\left(\mathbf{f}(\hat{\mathbf{x}})-\mathcal{K}\right)\cap \mathbf{f}(\Omega) = \mathbf{f}(\hat{\mathbf{x}})\label{MinSol}.
\end{align}
This definition of an optimal solution to \eqref{MOP} can be visualized geometrically, as in Figure \ref{fig:definition}. Here, in the case of $\mathcal{K}=\mathbb{R}_+^2$, the $\mathcal{K}$-optimal points are the points $\hat{\bf x}$ such that the cone $-\mathcal{K}$ routed at ${\bf f}(\hat{\bf x})$ is entirely outside the image of $\Omega$ under $\bf f$, denoted as $\bf f(\Omega)$ and also referred to as range of $\bf f$, apart from the unique point ${\bf f}(\hat{\bf x})$.  Furthermore, it can be observed that the set of optimal values is on the boundary of the set $\bf f(\Omega).$
We define $\mathcal{M}\left(\mathbf{f}\left(\Omega\right), \mathcal{K} \right)\subseteq\bf f(\Omega)$ as the set of all optimal values of \eqref{MOP} and $\mathcal{E}\left(\mathbf{f}\left(\Omega\right), \mathcal{K} \right)\subseteq\Omega$ as the set of all optimal points, called efficient set, i.e. the set of $\bf x\in\Omega$ achieving the values in $\mathcal{M}\left(\mathbf{f}\left(\Omega\right), \mathcal{K} \right)$. Moreover, the cone $\mathcal{K}  = \mathbb{R}_+^m$ induces the component-wise ordering on $\mathbb{R}^m$. In this case, $\mathcal{M}\left(\mathbf{f}\left(\Omega\right), \mathbb{R}_+^m\right)$ is referred to as the Pareto-Front or Pareto-Boundary, and any point ${\bf x} \in \mathcal{E}\left(\mathbf{f}\left(\Omega\right), \mathbb{R}_+^m\right)$ is referred to as Pareto-Optimal.
\subsection{Scalarization Approaches}
From Figure \ref{fig:definition} we can observe that a vector optimization problem usually has multiple $\mathcal{K}$-optimal solutions. In order to approximate the set of all $\mathcal{K}$-optimal solutions, the multi-objective problem is usually transformed into a single-objective problem, referred to as scalarization. Well known approaches are for example the Weighted-Sum-Method \cite{zadeh_weighted_sum}, the $\epsilon$-contraint method \cite{haimes1971bicriterion}, Chebyshev-scalarization, and the Pascoletti-Serafini scalarization \cite{pascoletti1984scalarizing}. In the following, the more general approach of Pascoletti-Serafini is presented. This approach will be considered as an alternative problem formulation to the usual weighted-sum-rate problem in this work.

\subsubsection{Pascoletti and Serafini Scalarization}
The approach by Pascoletti and Serafini \cite{pascoletti1984scalarizing} is stated as follows
\begin{subequations}\label{PS}
\begin{align}
	\min_{t,\mathbf{x}\in\Omega}&\, t\\
	\text{s.t.}&\, \mathbf{a}+t\mathbf{r}-\mathbf{f}(\mathbf{x}) \in \mathcal{K}.
\end{align}
\end{subequations}
for given parameters $\mathbf{a},\mathbf{r}$, called reference and direction vector, where $\mathbf{r}\in\mathcal{K}\setminus\lbrace0\rbrace$ and $\mathbf{a}\in\mathbb{R}^m$. This can be interpreted as moving the cone $-\mathcal{K}$ in the direction $\mathbf{r}$ on the line $\mathbf{a}+t\mathbf{r}$ until $(\mathbf{a}+t\mathbf{r}-\mathcal{K})\cap f(\Omega)$ is reduced to the empty set. Thus, the set of all minimal solutions can be obtained by changing $\mathbf{a},\mathbf{r}$ and solving \eqref{PS}.
Furthermore, the Weighted-Sum-Method can be interpreted as a special case of the Pascoletti and Serafini scalarization by using a weighted cone $\mathcal{K}_w = \lbrace \mathbf{x}\in\mathbb{R}^K \, | \, \mathbf{w}^T\mathbf{x}\geq 0 \rbrace$. The Pascoletti-Serafini scalarization with respect to a weighted cone can be formulated as
\begin{subequations}
    \label{eq:PS_weighted}
\begin{align}
	\min_{t,\mathbf{x}\in\Omega}&\, t\\
	\text{s.t.} &\, \mathbf{a}+t\mathbf{r}-\mathbf{f}(\mathbf{x}) \in \mathcal{K}_{\bf w}. 
\end{align} 
\end{subequations}
The following holds: $\hat{\mathbf{x}}$ is a solution of the weighted sum, i.e.,
\begin{align*}
    \min_{\mathbf{x}\in\Omega} \sum_{i=1}^n w_if_i(\mathbf{x})
\end{align*}
if and only if there exists a $\hat{t}$ s.t. $\left(\hat{t}, \hat{\mathbf{x}}\right)$ is a solution of \eqref{eq:PS_weighted}, see Theorem 2.39 in \cite{eichfelder_adaptive_2008}. Figure \ref{fig:example_weighted} illustrates the Pascoletti-Serafini Scalarization Approach for a weighted cone. The cone moves along the line ${\bf a} + t\bf r$ until the $\mathcal{K}_{\bf w}$-optimal point is found. Each set of weights yields a different weighted cone.

In general, the Pascoletti-Serafini Scalarization method can identify more points on the Pareto-Boundary compared to the Weighted-Sum-Method. As shown in Figure \eqref{fig:example_weighted}, the limitations of the Weighted-Sum-Method in parameterizing all points of the entire set $\mathcal{M}\left(\mathbf{f}\left(\Omega\right), \mathcal{K} \right)$ are evident. For instance, the point ${\bf f}\left(\hat{\bf x}\right)$ in Figure \ref{fig:definition} is not part of $\mathcal{M}\left(\mathbf{f}\left(\Omega\right), \mathcal{K}_{\bf w} \right)$ for any ${\bf w} \in \mathcal{K}^*$. However, if one is only interested to maximize or minimize the weighted sum, the weighted sum scalarization is adequate. Moreover, Pascoletti-Serafini Scalarization with a weighted cone is equivalent in this case, as shown by Theorem 2.39 in \cite{eichfelder_adaptive_2008}. 

In this work, we choose the formulation \eqref{eq:PS_weighted} instead of directly optimizing the weighted sum because \eqref{eq:PS_weighted} can be expressed in Lagrange form, leading to an iterative algorithm, which we are able to unfold in the sense of deep unfolding.

\begin{figure}[ht] 
     \centering 
     \begin{subfigure}{.5\textwidth} 
         \centering

\begin{tikzpicture}[x=0.75pt,y=0.75pt,yscale=-1,xscale=1]

\draw  [draw opacity=0][fill={rgb, 255:red, 229; green, 229; blue, 229 }  ,fill opacity=1 ] (435.6,107.8) -- (486,107.8) -- (486,151.8) -- (435.6,151.8) -- cycle ;
\draw  (342,223.7) -- (615.8,223.7)(369.38,21.2) -- (369.38,246.2) (608.8,218.7) -- (615.8,223.7) -- (608.8,228.7) (364.38,28.2) -- (369.38,21.2) -- (374.38,28.2)  ;
\draw  [line width=0.75] [line join = round][line cap = round] (429.42,56.36) .. controls (427.46,67.26) and (421.66,89.51) .. (433.3,98.23) .. controls (448.85,109.86) and (471.44,98.68) .. (485.28,108.62) .. controls (505.38,123.06) and (495.94,141.86) .. (515.14,154.24) .. controls (525.92,161.18) and (539.41,159.37) .. (546.21,149.04) .. controls (564.81,120.78) and (555.4,92.88) .. (533.36,69.64) .. controls (527.01,62.94) and (508.39,59.26) .. (502.3,57.81) .. controls (488.68,54.54) and (474.6,56.43) .. (460.78,55.5) .. controls (453.56,55.01) and (446.52,52.68) .. (439.28,52.61) .. controls (436.11,52.58) and (427.79,54.99) .. (430.62,56.36) ;
\draw    (486,107.8) -- (435.6,107.8) ;
\draw    (486,151.8) -- (486,107.8) ;
\draw  [fill={rgb, 255:red, 0; green, 0; blue, 0 }  ,fill opacity=1 ] (483.3,107.8) .. controls (483.3,106.31) and (484.51,105.1) .. (486,105.1) .. controls (487.49,105.1) and (488.7,106.31) .. (488.7,107.8) .. controls (488.7,109.29) and (487.49,110.5) .. (486,110.5) .. controls (484.51,110.5) and (483.3,109.29) .. (483.3,107.8) -- cycle ;

\draw (513.6,97.8) node [anchor=north west][inner sep=0.75pt]  [font=\scriptsize]  {${\bf f}( \Omega )$};
\draw (449.4,126.2) node [anchor=north west][inner sep=0.75pt]  [font=\scriptsize]  {$-\ \mathcal{K}$};
\draw (474.2,87.8) node [anchor=north west][inner sep=0.75pt]  [font=\scriptsize]  {${\bf f}(\hat{\bf x})$};
\draw (620,216.6) node [anchor=north west][inner sep=0.75pt]  [font=\scriptsize]  {$f_{1}$};
\draw (361.2,1.4) node [anchor=north west][inner sep=0.75pt]  [font=\scriptsize]  {$f_{2}$};

\end{tikzpicture}

         \caption{A $\mathcal{K}$-optimal point $\mathbf{f}(\hat{\mathbf{x}})$.}  \label{fig:definition}
     \end{subfigure}\\
     \begin{subfigure}{.5\textwidth} \centering    
     \begin{tikzpicture}[x=0.75pt,y=0.75pt,yscale=-1,xscale=1]

\draw  [draw opacity=0][fill={rgb, 255:red, 229; green, 229; blue, 229 }  ,fill opacity=1 ][line width=0.75]  (85.84,75.81) -- (30.2,89.56) -- (157.76,213.4) -- (213.4,199.64) -- cycle ;
\draw  (19.6,224.5) -- (293.4,224.5)(46.98,22) -- (46.98,247) (286.4,219.5) -- (293.4,224.5) -- (286.4,229.5) (41.98,29) -- (46.98,22) -- (51.98,29)  ;
\draw  [line width=0.75] [line join = round][line cap = round] (106.22,57.16) .. controls (104.26,68.06) and (98.46,90.31) .. (110.1,99.03) .. controls (125.65,110.66) and (148.24,99.48) .. (162.08,109.42) .. controls (182.18,123.86) and (172.74,142.66) .. (191.94,155.04) .. controls (202.72,161.98) and (216.21,160.17) .. (223.01,149.84) .. controls (241.61,121.58) and (232.2,93.68) .. (210.16,70.44) .. controls (203.81,63.74) and (185.19,60.06) .. (179.1,58.61) .. controls (165.48,55.34) and (151.4,57.23) .. (137.58,56.3) .. controls (130.36,55.81) and (123.32,53.48) .. (116.08,53.41) .. controls (112.91,53.38) and (104.59,55.79) .. (107.42,57.16) ;
\draw    (85.86,75.8) -- (213.4,199.64) ;
\draw  [fill={rgb, 255:red, 0; green, 0; blue, 0 }  ,fill opacity=1 ] (99.4,159.1) .. controls (99.4,157.61) and (100.61,156.4) .. (102.1,156.4) .. controls (103.59,156.4) and (104.8,157.61) .. (104.8,159.1) .. controls (104.8,160.59) and (103.59,161.8) .. (102.1,161.8) .. controls (100.61,161.8) and (99.4,160.59) .. (99.4,159.1) -- cycle ;
\draw    (102.1,159.1) -- (120.4,140.43) ;
\draw [shift={(121.8,139)}, rotate = 134.42] [color={rgb, 255:red, 0; green, 0; blue, 0 }  ][line width=0.75]    (10.93,-3.29) .. controls (6.95,-1.4) and (3.31,-0.3) .. (0,0) .. controls (3.31,0.3) and (6.95,1.4) .. (10.93,3.29)   ;
\draw    (102.1,159.1) -- (219.61,38.04) ;
\draw [shift={(221,36.6)}, rotate = 134.15] [color={rgb, 255:red, 0; green, 0; blue, 0 }  ][line width=0.75]    (10.93,-3.29) .. controls (6.95,-1.4) and (3.31,-0.3) .. (0,0) .. controls (3.31,0.3) and (6.95,1.4) .. (10.93,3.29)   ;
\draw  [fill={rgb, 255:red, 0; green, 0; blue, 0 }  ,fill opacity=1 ] (106.2,97.5) .. controls (106.2,96.01) and (107.41,94.8) .. (108.9,94.8) .. controls (110.39,94.8) and (111.6,96.01) .. (111.6,97.5) .. controls (111.6,98.99) and (110.39,100.2) .. (108.9,100.2) .. controls (107.41,100.2) and (106.2,98.99) .. (106.2,97.5) -- cycle ;

\draw (295.6,216) node [anchor=north west][inner sep=0.75pt]  [font=\scriptsize]  {$f_{1}$};
\draw (190.4,98.6) node [anchor=north west][inner sep=0.75pt]  [font=\scriptsize]  {${\bf f}( \Omega )$};
\draw (89.4,156.2) node [anchor=north west][inner sep=0.75pt]  [font=\scriptsize]  {$\bf a$};
\draw (121.8,139) node [anchor=north west][inner sep=0.75pt]  [font=\scriptsize]  {$\bf r$};
\draw (222.4,26.4) node [anchor=north west][inner sep=0.75pt]  [font=\scriptsize]  {${\bf a}\ +\ t{\bf r}\ -\ \mathcal{K}_{w}$};
\draw (112,81.4) node [anchor=north west][inner sep=0.75pt]  [font=\scriptsize]  {${\bf f}(\hat{\bf x})$};
\draw (156,182.2) node [anchor=north west][inner sep=0.75pt]  [font=\scriptsize]  {$-\ \mathcal{K}_{w}$};
\draw (39.2,3.2) node [anchor=north west][inner sep=0.75pt]  [font=\scriptsize]  {$f_{2}$};

\end{tikzpicture}

\caption{Moving a weighted cone $\mathcal{K}_{\bf w}$ along the line ${\bf a} + t\bf r$.}\label{fig:example_weighted}
         \end{subfigure} 
     \caption{Vector optimization: a) definition of a $\mathcal{K}$-optimal point with $\mathcal{K}=\mathbb{R}_+^2$ and b) an example of moving a weighted cone along $\mathbf{a}+\mathbf{r}$.}
\end{figure}
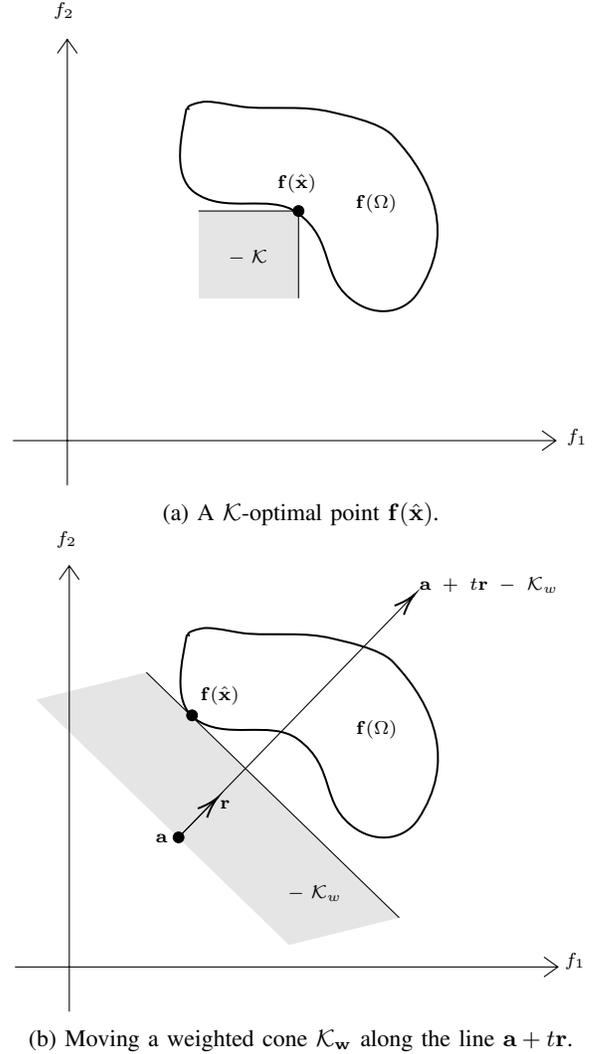 
\begin{figure}
    \centering
    
\begin{tikzpicture}[x=0.75pt,y=0.75pt,yscale=-1,xscale=1]

\draw  [draw opacity=0][fill={rgb, 255:red, 229; green, 229; blue, 229 }  ,fill opacity=1 ][line width=0.75]  (302.79,55.12) -- (167,55.12) -- (325.21,182.12) -- (461,182.12) -- cycle ;
\draw  (466.32,22.7) -- (192.52,22.7)(438.94,225.2) -- (438.94,0.2) (199.52,27.7) -- (192.52,22.7) -- (199.52,17.7) (443.94,218.2) -- (438.94,225.2) -- (433.94,218.2)  ;
\draw    (438.52,202.8) .. controls (418.52,102.4) and (352.52,38.8) .. (226.52,22.8) ;
\draw  [dash pattern={on 0.84pt off 2.51pt}]  (226.12,239.2) -- (226.52,22.8) ;
\draw  [dash pattern={on 0.84pt off 2.51pt}]  (438.52,202.8) -- (190.52,202.8) ;
\draw  [fill={rgb, 255:red, 0; green, 0; blue, 0 }  ,fill opacity=1 ] (228.12,202.9) .. controls (228.12,204.39) and (226.91,205.6) .. (225.42,205.6) .. controls (223.92,205.6) and (222.72,204.39) .. (222.72,202.9) .. controls (222.72,201.41) and (223.92,200.2) .. (225.42,200.2) .. controls (226.91,200.2) and (228.12,201.41) .. (228.12,202.9) -- cycle ;
\draw    (225.42,202.9) -- (437.41,23.99) ;
\draw [shift={(438.94,22.7)}, rotate = 139.84] [color={rgb, 255:red, 0; green, 0; blue, 0 }  ][line width=0.75]    (10.93,-3.29) .. controls (6.95,-1.4) and (3.31,-0.3) .. (0,0) .. controls (3.31,0.3) and (6.95,1.4) .. (10.93,3.29)   ;
\draw    (225.42,202.9) -- (253.38,179.68) ;
\draw [shift={(254.92,178.4)}, rotate = 140.29] [color={rgb, 255:red, 0; green, 0; blue, 0 }  ][line width=0.75]    (10.93,-3.29) .. controls (6.95,-1.4) and (3.31,-0.3) .. (0,0) .. controls (3.31,0.3) and (6.95,1.4) .. (10.93,3.29)   ;
\draw    (461,182.12) -- (302.78,55.12) ;

\draw (167.32,13.8) node [anchor=north west][inner sep=0.75pt]  [font=\scriptsize]  {$-R_{1}$};
\draw (426.12,226) node [anchor=north west][inner sep=0.75pt]  [font=\scriptsize]  {$-R_{2}$};
\draw (358.92,32.6) node [anchor=north west][inner sep=0.75pt]  [font=\scriptsize]  {$-{\bf R}( \Omega )$};
\draw (227.,204.2) node [anchor=north west][inner sep=0.75pt]  [font=\scriptsize]  {${\bf u}\ =\ {\bf a}$};
\draw (233.32,167.6) node [anchor=north west][inner sep=0.75pt]  [font=\scriptsize]  {$\bf r$};
\draw (368.73,138.47) node [anchor=north west][inner sep=0.75pt]  [font=\scriptsize]  {$-\mathcal{K}_{w}$};
\draw (271.2,87.2) node [anchor=north west][inner sep=0.75pt]  [font=\scriptsize]  {${\bf a}\ +\ t{\bf r}\ -\ \mathcal{K}_{w}$};

\end{tikzpicture}

    \caption{Example for Pascoletti and Serafini scalarization, with $\bf a,r$ based on the Utopian point, for weighted cone $\mathcal{K}_{\bf w}$. Here for $K=2$.}
    \label{fig:example_log}
\end{figure}
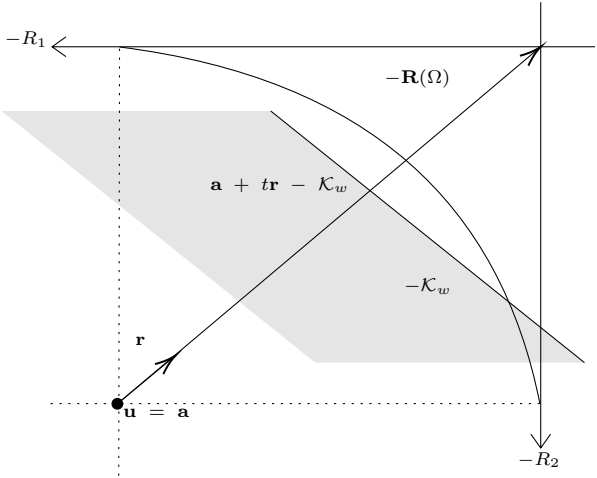


\section{Proposed Algorithm}\label{sec:main}
In the following, we first formulate the weighted-sum-rate maximization problem in \eqref{eq:weightedSum} in terms of the equivalent Pascoletti-Serafini scalarization. Then, we derive a primal-dual gradient algorithm to approximate the solution of this new formulation. This is followed by an introduction to deep unfolding, which interprets each iteration of the proposed primal-dual algorithm as layers of a neural network. Eventually, this yields the unfolded algorithm proposed in this work, whose parameters can be optimized in a data-driven unsupervised way.

\subsection{Primal-Dual Algorithm}\label{sec:main_pda}
The tuple of rates for each link can be considered as a vector-valued function, i.e.
\begin{align}
\mathbf{R}(\mathbf{x}, \mathbf{G}) &= (R_1(\mathbf{x}, \mathbf{G}),R_2(\mathbf{x}, \mathbf{G}),\dots, R_K(\mathbf{x}, \mathbf{G}))\in\mathbb{R}^K \label{eq:rate_vector_function}
\end{align}
and thus we consider
\begin{align}
\max_{\mathbf{x}\in [ 0,1]^K}\mathbf{R}(\mathbf{x}, \mathbf{G}) \text{ w.r.t. }\mathcal{K}_{\bf w}\label{eq:Problem}.
\end{align} 
As discussed, an alternate scalarization approach for \eqref{eq:Problem} can then be stated as
\begin{subequations}\label{eq:scala}
\begin{align}
&\min_{t\in\mathbb{R}, \mathbf{x}\in [0,1]^K} t  \\
\text{s.t.}&\, {\bf w}^T\left(\mathbf{a}+t\mathbf{r}+\mathbf{R}(\mathbf{x}, \mathbf{G}) \right)\geq 0. 
\end{align}
\end{subequations}
Since the enumeration of links is arbitrary and the original problem \eqref{eq:weightedSum} is permutation invariant, we chose the reference and direction vector in \eqref{eq:scala} w.r.t. the Utopian point. The Utopian point is defined as the point with components $\bf{u}_i$ equal to the absolute maxima of each component of the function $R_i$, the Utopian point lies generally outside of the range of the multi-objective function under it's domain, and is therefore non achievable. In our case, this is given by letting 
\begin{align}
    {u}_i = \max_{\mathbf{x}\in [ 0,1]^K}{R}_i(\mathbf{x}, \mathbf{G}) ={R}_i(\mathbf{e}_i, \mathbf{G})\, \forall\, i,\label{eq:Utopian}
\end{align}
where ${\bf e}_i$ is the unit vector, since ${ x}_i = 1$ and ${ x}_j = 0$, for all $j \neq i$ yields that only Tx $i$ is active and therefore the rate $R_i$ of link $i$ is maximized, since the desired signal is transmitted at maximum power and there is no interference form the other links. In the following we define the reference and direction vector as ${\bf a} = -{\bf u}$ and ${\bf r} = \bf u/\|u\|_2$. This aligns with results in \cite{khaledian2016restricting}, here it was shown that the parameter set can be restricted to ${\bf r}\in \text{int}\mathcal{K} \cap \lbrace {\bf x}: \|x\|_2 = 1 \rbrace$ if the objective is shifted s.t. the Utopian point is the origin, which can also be done here by setting $\bf a = -\bf u$. By doing so, the cone defined by the weights will be moved along the line starting from the Utopian point into the origin. Furthermore with this choice of reference and direction vector the objective in \eqref{eq:scala} is bounded from below, see Theorem 4.2. in \cite{khaledian2016restricting}. Additionally \eqref{eq:scala} becomes permutation invariant w.r.t. the enumeration of links. An example of the constructed problem is given in Figure \ref{fig:example_log}.

In order to derive an iterative algorithm from \eqref{eq:scala} the corresponding Lagrangian problem and a classical primal-dual gradient algorithm is developed.
The Lagrangian function and it's corresponding derivatives can be written as
\begin{align}
\mathcal{L}\left(t,\mathbf{x},\lambda\right) & = t - \lambda{\bf w}^T\left(\mathbf{a} + t\mathbf{r} + \mathbf{R}(\mathbf{x}, \mathbf{G}) \right)\label{eq:lagrangian} \\ 
\partial_t \mathcal{L}\left(t,\mathbf{x},\lambda\right) & = 1 - \lambda{\bf w}^T \mathbf{r} \nonumber\\
\nabla_\mathbf{x} \mathcal{L}\left(t,\mathbf{x},\lambda\right) & = - \lambda J(\mathbf{R}(\mathbf{x}, \mathbf{G}))^T{\bf w} \nonumber \\
\partial_\lambda \mathcal{L}\left(t,\mathbf{x},\lambda\right) & = -{\bf w}^T\left(\mathbf{a} + t\mathbf{r} + \mathbf{R}(\mathbf{x}, \mathbf{G})\right), \nonumber
\end{align}
with $\lambda \geq 0,$ and where $J(\mathbf{R}(\mathbf{x}, \mathbf{G}))$ is the Jacobian matrix of $\mathbf{R}(\mathbf{x}, \mathbf{G})$ w.r.t. $\mathbf{x}\in [0,1]^K$. From here we can derive the  primal dual algorithm, by using gradient descent to update $t$, projected gradient descent\footnote{Gradient descent since we derived the Pascoletti-Serafini scalarization for $\min$ and thus we have to reformulate \eqref{eq:Problem} into a minimization problem.} to update $\mathbf{x}$ and a gradient ascent to update the Lagrangian $\lambda$, to approximate a stationary point of the Lagrangian function \eqref{eq:lagrangian}. Overall the primal-dual algorithm is stated as
\begin{align*}
 	t^{(k)} &=t^{(k-1)}-\alpha_1 \left(1-\lambda^{(k-1)}{\bf w}^T\mathbf{r}\right)\\
	\mathbf{x}^{(k)} &= \mathcal{P}_{[0,1]}\left(\mathbf{x}^{(k-1)}+\tilde{\alpha_2} \lambda^{(k-1)}J(\mathbf{R}(\mathbf{x}^{(k-1)}, \mathbf{G}))^T{\bf w}\right)\\
	\lambda^{(k)} & = \mathcal{P}_{\geq 0}\left(\lambda^{(k-1)}-\alpha_3 {\bf w}^T\left(\mathbf{a}+t^{(k)}\mathbf{r}+\mathbf{R}(\mathbf{x}^{(k)}, \mathbf{G})\right)\right).
\end{align*}
Here $\mathcal{P}_{[0,1]}$ is the component wise projection on the interval $[0,1]$. Moreover, a projection $\mathcal{P}_{\geq 0}(x) = \max\lbrace x, 0\rbrace$ to the update of $\mathbf{\lambda}$ is added to ensure $\lambda\geq 0$. Furthermore, a normalized gradient update for $\bf x$ as $\tilde{\alpha}_2 = \alpha_2/\|J({\bf R}({\bf x}^{(k-1)}, {\bf G}))^T{\bf w}\|_2$ is introduced. It could be observed in experiments that this \textit{stabilizes} the behaviour of the objective \eqref{eq:weightedSum} w.r.t. the iterations, see Section \ref{sec:num_untrained}. These final additions to the primal-dual algorithm are also incorporated to facilitate easier training of the unfolded algorithm. 


\subsection{Deep Unfolding}
Deep unfolding involves viewing the iterations of an iterative algorithm as layers of a neural network, and then optimizing the algortihms parameters through training, i.e. using a stochastic gradient to minimize a loss function, rather than empirically adjusting them. To achieve this, the iterations are fixed and executed sequentially as stages of multilayer computation. Consequently, certain parameters in each layer are replaced as learnable parameters during the training process. This approach has  already been succesfully applied to a variety of iterative algorithms, e.g. learned iterative shrinkage thresholding algorithm \cite{liu2019alista, chen2018convlista, hauffen2022algorithm}, learned alternate message passing algorithm \cite{7934066, osman_AMP_2021}, or learned alternating direction of multipliers \cite{8550778, miriya2022deep}.

One can formalize the general idea of deep unfolding as follows. Define the following update rule
\begin{align*}
	T_{\Theta^{(k)}}\left( \cdot\, ;\, y\right):= T\left( \cdot\, ;\,\Theta^{(k)},\, y\right),
\end{align*}
where $T\left( \cdot\, ;\,\Theta,\, y\right)$ is an iteration of some iterative algorithm, for example a gradient descent step, w.r.t. an input $y$ and parameters $\Theta$, e.g. a step-size. Here, $\Theta^{(k)}$ are the trainable parameters for layer/iteration $k$ and $\Theta = \cup_{k=1}^N \Theta^{(k)}$ are the trainable parameters for all layers/iterations, i.e. the whole network. The neural network/learned algorithm is defined as the concatenation of the $N$ layers
\begin{align*}
	\left( T_{\Theta^{(N)}} \circ \dots \circ T_{\Theta^{(1)}} \right)  \left(x^{(0)}\, ; y\right) = x^{(N)}\,.
\end{align*}
The so defined neural network can then be trained via supervised or unsupervised learning, i.e. by data-driven optimization, with a standard optimizer w.r.t. $\Theta = \bigcup_{k=1}^N \Theta^{(k)}$, e.g. the widely used Adam Optimizer \cite{kingma2014adam}. 

In contrast to deep neural networks the so defined network is usually trained layer-wise and not end-to-end. In end-to-end training one  optimizes the network parameters with some loss function w.r.t. the output of the network. While the unfolded algorithm is still an iterative procedure, the output of each layer always has the same dimension and thus one can optimize it's parameters layer-wise, \cite{chen2018convlista, chen_hyperparameter_2021, osman_AMP_2021}. In layer-wise training one iteratively takes the output of each layer and optimizes the parameters only of this current layer. Usually this is followed by refinement steps, where still the output of the current layer is taken but now all already seen trainable parameters are optimized, with a finer training-rate. This procedure is done iteratively, going through all layers of the network, hence the name layer-wise training. In Section \ref{sec:main_lpda} this is explained in more detail for the proposed algorithm.  

\subsection{Learned Primal Dual Algorithm}\label{sec:main_lpda}
Building upon the previously described scalarization approach and the subsequent discussion, we are now able to propose the learned primal-dual algorithm for power control, as outlined in Algorithm \ref{LearnedPD}.
\begin{algorithm}
\textbf{Input: }$\mathbf{G}, \bf w$, $\mathbf{x}^{(0)}$ \\
 \For{$k = 1,\dots, N$}{
 	$t^{(k)} = \, t^{(k-1)}-\alpha_1^{(k)}(1-\lambda^{(k-1)}{\bf w}^T\mathbf{r})$\\
	$\mathbf{x}^{(k)} = \mathcal{P}_{[0,1]}\left(\mathbf{x}^{(k-1)}+\tilde{\alpha_2}^{(k)}\lambda^{(k-1)}J(\mathbf{R}(\mathbf{x}^{(k-1)}, \mathbf{G}))^T{\bf w}\right)$\\
    $\tilde{\alpha}_2^{(k)} = \alpha_2^{(k)}/\|J({\bf R}({\bf x}^{(k-1)}, {\bf G}))^T{\bf w}\|_2$\\
	$\lambda^{(k)} = \mathcal{P}_{\geq 0}\left(\lambda^{(k-1)}-\alpha_3^{(k)} {\bf w}^T\left(\mathbf{a}+t^{(k)}\mathbf{r}+\mathbf{R}(\mathbf{x}^{(k)}, \mathbf{G})\right)\right)$
 }
\caption{Iterative Vector Algorithm}\label{LearnedPD}
\end{algorithm}
The trainable parameters are defined as $\Theta = \bigcup_{k=1}^N\Theta^{(k)},$ where $\Theta^{(k)} = \left\lbrace \alpha_1^{(k-1)}, \alpha_2^{(k)} , \alpha^{(k-1)}_3\right\rbrace$, and $N$ is the number of layers. Notably, only the step sizes that influence the current iterate $k$ are included in the trainable parameters $\Theta^{(k)}$ of layer/iteration $k$. Therefore, we incorporate $\alpha_1^{(k-1)},  \alpha^{(k-1)}_3$, but not $\alpha_1^{(k)},  \alpha^{(k)}_3$ since the latter do not have an immediate effect on the current iterate ${\bf x}^{(k)}$, causing their gradients to vanish. On the other hand $\alpha_1^{(k-1)},  \alpha^{(k-1)}_3$ have an indirect influence on ${\bf x}^{(k)}$, since they have a direct influence on $\lambda^{(k-1)}$. Consequently, if the trainable parameters for each layer consisted only of the current step sizes $\alpha_i^{(k)}$, then only $\alpha_2^{(k)}$ would be optimized during the first training stage, as the gradient for the other step sizes would vanish. 

In general, there are a total of $3N$ trainable parameters for the proposed algorithm with $N$ layers/iterations.
\subsubsection{Training Procedure}\label{sec:main_train}
As previously discussed, we train the primal-dual algorithm in an unsupervised manner and employ layer-wise training. This means that the layer are trained iteratively instead of training all layers at once end-to-end. Each layer is trained in two stages. While the first stage optimizes only the trainable parameters $\Theta^{(k)}$ of the current layer $k$, the second stages optimizes the layer w.r.t. all trainable parameters $\Theta$ using a smaller training rate. The second stage is usually referred to as the refinement stage and is repeated a few times with a decreasing training rate. 

The first stage in training layer $k$, starts with optimizing only $\Theta^{(k)}$ based on a cost function of ${\bf x}^{(k)}$. After this stage is done, layer $k$ will be refined by optimizing w.r.t. all parameters $\Theta$, using a smaller training rate, and again a cost function of ${\bf x}^{(k)}$. Once the final layer has been trained and refined, the training process is complete. 

The training for each layer can be stated as following. The first stage aims to optimize
\begin{align*}
\min_{\Theta^{(k)}} \mathbb{E}_{\mathbf{G}\text{\raisebox{-0.9ex}{\~{}}}\mathcal{D}_1, \mathbf{w}\text{\raisebox{-0.9ex}{\~{}}}\mathcal{D}_2  }\left[\ell\left(\mathbf{x}^{(k)}\left(\mathbf{G}, \Theta^{(k)}\right), \mathbf{w}\right)\right],
\end{align*}
here $\mathcal{D}_i$ are distributions, where the weights and system matrices are drawn from. For example we will set $\mathcal{D}_2 = \mathcal{U}_{[0,1]}$, generating samples from $\mathcal{D}_1$ will be discussed in Section \ref{sec:num_exp}. 
As said before, the second stage, i.e. refinement stage, still trains the current iteration and optimizes the loss function w.r.t. all trainable parameters. This can be formulated as
\begin{align*}
\min_{\Theta }\mathbb{E}_{\mathbf{G}\text{\raisebox{-0.9ex}{\~{}}}\mathcal{D}_1, \mathbf{w}\text{\raisebox{-0.9ex}{\~{}}}\mathcal{D}_2  }\left[\ell\left(\mathbf{x}^{(k)}\left(\mathbf{G}, \Theta\right), \mathbf{w}\right)\right].
\end{align*}
The loss function is given as
\begin{align*}
\ell\left(\mathbf{x}^{(k)}\left(\mathbf{G}, \Theta\right), \mathbf{w}\right) = & -\sum_{i=1}^Kw_iR_i\left(\mathbf{x}^{(k)}\left(\mathbf{G}, \Theta\right)\right) 
\end{align*}
where $\mathbf{x}^{(k)}$ is the output of the $k$-th iteration depending on $\mathbf{G}$ and the parameters $\Theta$. 
\begin{algorithm}
    \textbf{Input: }$\Theta =\bigcup_{k=1}^N\Theta^{(k)}, t_r, f_1, f_2$\\
    \For{$k=1,\dots, N$}
   {
   \For{t $=1,\dots,$ \text{MaxIter}}{
    Generate $\lbrace\mathbf{G}_b, \mathbf{w}_b\rbrace_{b = 1}^{n_\text{train}}$\\
    Compute $\ell\left(\mathbf{x}^{(k)}\left(\mathbf{G}, \Theta^{(k)}\right), \mathbf{w}\right)$\\
    AdamOptimizer($t_r$).min\big($\ell$, var list=$\Theta^{(k)}$\big)
    }
   \For{$f \in\lbrace f_1, f_2\rbrace$}{
   \For{t $=1,\dots,$ \text{MaxIter}}{
    Generate $\lbrace\mathbf{G}_b, \mathbf{w}_b\rbrace_{b = 1}^{n_\text{train}}$\\
    Compute $\ell\left(\mathbf{x}^{(k)}\left(\mathbf{G}, \Theta\right), \mathbf{w}\right)$\\
    AdamOptimizer($t_r\cdot f$).min\big($\ell$, var list=$\Theta$\big)
    }}
    
   }
     \caption{Training}\label{alg:training}
\end{algorithm}

The expected value in the latter equations is in practive approximated by sampling from the distributions $\mathcal{D}_1,\mathcal{D}_2$. In each training iteration, a set of $n_\text{train}$ training D2D networks are generated, i.e. matrices $\mathbf{G}$ and vectors $\mathbf{w}$, which are used to update the trainable parameters. The loss function is evaluated on a validation set and the training is stopped for the current stage when the loss converges or the number of maximum iterations ${MaxIter}$ is reached. 

A pseudo code for training is shown in Algorithm \ref{alg:training}, for the sake of readability, the stopping criteria is not included here. Observe the two stages of training a single layer: the first loop optimizes only with respect to $\Theta^{(k)}$, where in the second loop, all trainable parameters $\Theta$ are optimized. The \textit{AdamOptimizer} refers to the widely recognized stochastic gradient descent algorithm extensively used in deep learning applications \cite{kingma2014adam}. In Algorithm \ref{alg:training}, we adopt TensorFlow syntax. Specifically, the expression AdamOptimizer($t_r$).min\big($\ell$, var list=$\Theta$\big) denotes a single iteration of the Adam algorithm with a step size or learning rate $t_r$ to minimize the loss function $\ell$ w.r.t. the trainable parameters $\Theta$.

In the following we will refer to Algorithm \ref{LearnedPD} as Vanilla Iterative Vector Algorithm (VIVA) when referreing to the algorithm without trained parameters, and as Learned Unfolded Vector Algorithm (LUVA) when referring to the algoritm with trained parameters.
\subsubsection{Complexity} The highest complexity in Algorithm \ref{LearnedPD} lies in the updates of $\bf x$ in line 4 and in evaluating the instantaneous rates ${\bf R}$ in line 6 of Algorithm \ref{LearnedPD}, which can be written as
\begin{align}
    { R}_i({\bf x}) = \log\left(\sigma^2 + ({\bf Gx})_i\right) - \log\left(\sigma^2 + ({\bf \tilde{G}x})_i\right) \label{eq:dc_rep}
\end{align}
where $\bf \tilde{G}$ is $\bf G$ with diagonal equal to zero, i.e. $\tilde{\bf G} = {\bf G} - \text{diag}({\bf G}_{ii})$. Evaluating the rates has thus a complexity of $\mathcal{O}\left(K^2\right)$, due to the matrix-vector multiplication. Evaluating the Jacobian can be done using the already computed matrix vector products in \eqref{eq:dc_rep} (which is done in the previous iteration), since the Jacobian is given as
\begin{align*}
    J(\mathbf{R}(\mathbf{x}^{(k)}, \mathbf{G}))_{ij} = \frac{{\bf G}_{ij}}{\sigma^2 + ({\bf Gx})_i} - \frac{{\bf \tilde{G}}_{ij}}{\sigma^2 + ({\bf \tilde{G}x})_i}.
\end{align*}
Therefore, the proposed algorithm, similar to any algorithm which needs to evaluate the gradient or Jacobian of \eqref{eq:weightedSum}, has a complexity of $\mathcal{O}\left(K^2\right)$ per iteration. 

Note that learning is carried out offline, while Algorithm \ref{LearnedPD} (with learned parameters) is applied at run-time. Therefore, the complexity for the real-time operations of power control is only the one of Algorithm \ref{LearnedPD} while the training has nothing to do with the run-time complexity of the proposed algorithm.


\section{Numerical Experiments}\label{sec:num_exp}
For the following numerical experiments, the distance-dependent path loss model outlined in ITU-1411 is considered. Specifically, $K$ transmitters, denoted as $\text{Tx}_i$, are uniformly distributed within a square area of length $500$m. The corresponding receivers, $\text{Rx}_i$, are positioned within a disk centered around each transmitter, with a distance uniformly chosen between $d_\text{min}$ to $d_\text{max}$. The respective system parameters can be found in Table \ref{tab:system_params}. For simplicity the same transmit power for each transmitter is used, i.e. $P_i = P$ for all $i$. In the following we are going to present observations from the untrained algorithm, i.e. VIVA, which influenced some already mentioned design decision. We conclude with experiments on the trained unfolded Algorithm \ref{LearnedPD}, i.e. LUVA. Overall it is presented that LUVA is able to achieve almost always the same performance as FPLinQ, with a lower number of iterations and thus a lower complexity. Moreover we experimentally show strong generalizability of the trained unfolded algorithm, by testing LUVA on different scenarios, for example changing the number of users or signal-to-interferences ratio (SNR) levels. 

The proposed algorithm's ability to operate effectively on arbitrary weights demonstrates its suitability as a building block in a dynamic fairness scheduling scheme that maximizes a suitable component-wise increasing concave network utility function of the long-term average per-link throughput rates, via the general method of virtual queues and Lyapunov Drift Plus Penalty (LDPP), \cite{neely2010stochastic, georgiadis2006resource, neely2012stability}.
\begin{table}[]
    \centering
    \begin{tabular}{c|c}
        Bandwidth & $20$ MHz \\ \hline
        Carrier frequency & $2.4$ GHz\\ \hline
        Antenna height & $1.5$ m\\ \hline
        Transmit power level & $20$ dBm\\ \hline
        Background noise level & $-174$ dBm/Hz \\ \hline
        $d_\text{min} \text{ \raisebox{-0.9ex}{\~{}} } d_\text{max}$ & $2 \text{ m}\text{ \raisebox{-0.9ex}{\~{}} } 65$ m 
    \end{tabular}
    \caption{System Parameters for ITU-1411 short-range outdoor model used in the following numerical experiments.}
    \label{tab:system_params}
\end{table}
\subsection{Performance of the Untrained Algorithm}\label{sec:num_untrained}
Before presenting the results of LUVA we investigate the behaviour of the untrained algorithm, i.e. VIVA. It can be observed that VIVA is very sensitive w.r.t. the choice of parameters. 

\begin{figure}[htp]
    \centering 
    \begin{subfigure}{1.\linewidth}
        \centering
        \includegraphics[width=0.9\textwidth]{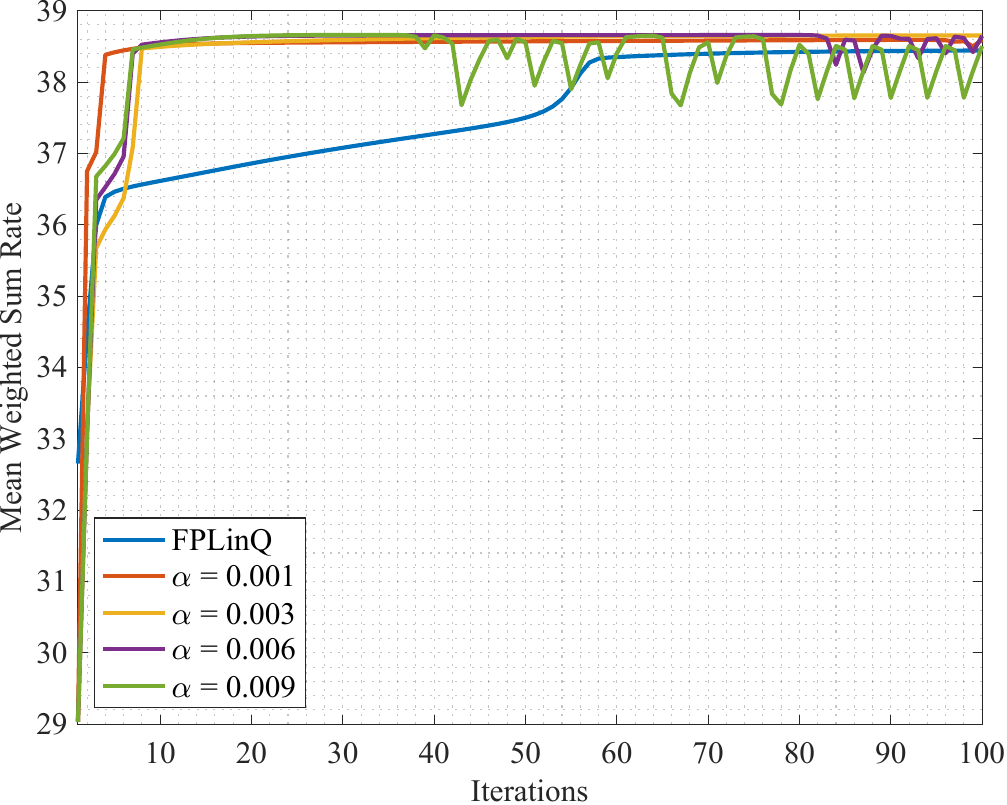}
        \caption{Primal-dual algorithm, without modifications discussed in Section \ref{sec:main_pda}.}
        \label{fig:osc}
    \end{subfigure}
    \vskip\baselineskip
    \begin{subfigure}{1.\linewidth}
        \centering
        \includegraphics[width=0.9\textwidth]{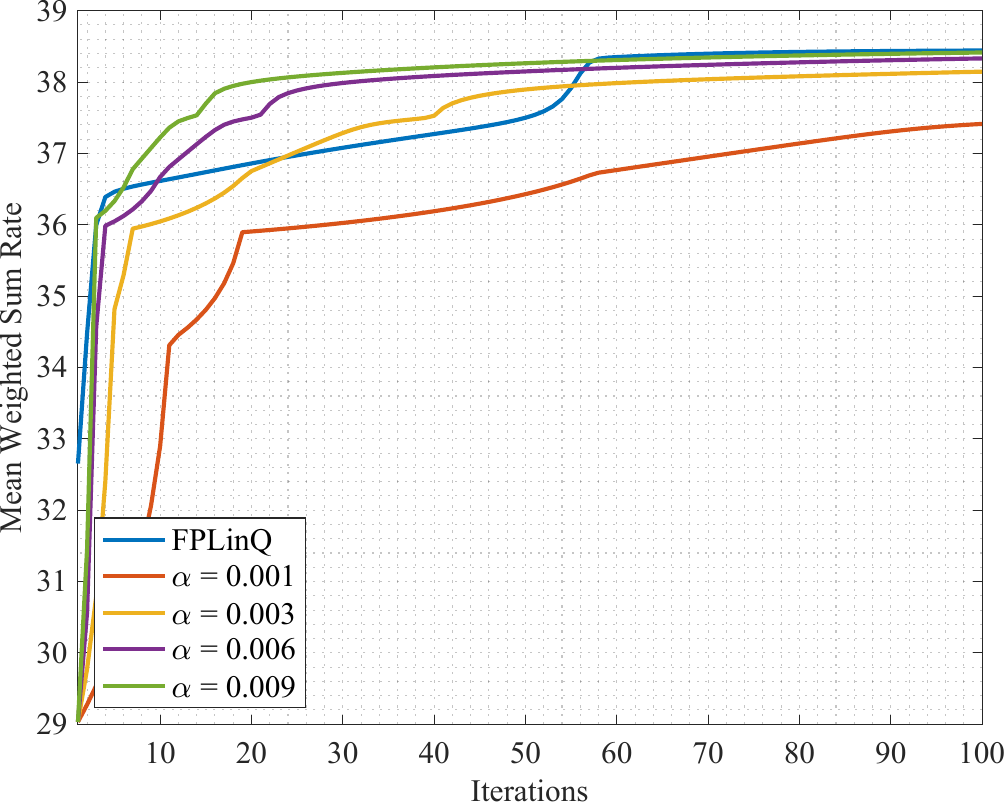}
        \caption{Primal-dual algorithm with Normalization and projection for $\lambda$, as discussed in Section \ref{sec:main_pda}}
        \label{fig:osc_1}
    \end{subfigure}
    \caption{Impact of normalization and enforcing constraints on $\lambda$, by testing the algorithm on the same network, with different step-sizes: a) The primal-dual algorithm, without normalization and projection of $\lambda$. b) primal-dual algorithm with discussed normalization and projection. It can be observed that in a) a slight change in step-size already leads to oscillation, while in b) this is not the case.}
    \label{fig:justifying_normalization}
\end{figure}

See for example Figure \ref{fig:osc}, here the \textit{standard} primal dual algorithm, i.e. without normalizing the gradient in the update for $\bf x$ and without forcing $\lambda\geq 0$, is shown for one D2D-network, generated as discussed. In Figure \ref{fig:osc_1} the modified version is presented, i.e. with introducing the normalizing factor in the updates for $\bf x$ and forcing $\lambda$ to be non-negative. 

In both Figures the step-sizes are changed from $\alpha_i\in\lbrace 0.001, 0.003,0.006, 0.009\rbrace$. Small changes already cause a fluctuating behavior in the weighted-sum-rate, while by adding the normalization and projection of ${\lambda}$ this behavior can be absorbed. Moreover it can already be observed that the proposed algorithm and alternative scalarization approach is able to outperform FPLinQ in this case.

This serves only as an example. Fine-tuning the step-sizes for gradient updates without normalization and ensuring $\lambda\geq 0$ would yield also a better and non-oscillating performance. However, this would result in different set of step-sizes for different D2D-networks, which is not practical here. Moreover by data-driven optimization of the parameters we optimize the expected value of instantaneous rates over a distribution of D2D-networks and random weights instead for one network, by normalizing the gradients as discussed and applying deep unfolding we are able to find step-sizes not only for one layout but for a distribution of D2D-networks, as well as increasing the convergence speed of the proposed algorithm, which is presented in the coming sections. Nonetheless we can already see that with fine-tuning the parameters we are able to outperform FPLinQ. 

\subsection{Performance of the Unfolded Algorithm}
The performance of the proposed network is investigated with respect to FPLinQ \cite{shen_fplinq_2017, shen_fractional_2018-1}. The two methods are compared w.r.t. the following performance metric
\begin{align}
\mathbb{E}_{\mathbf{G}\text{\raisebox{-0.9ex}{\~{}}}\mathcal{D}_1, \mathbf{w}\text{\raisebox{-0.9ex}{\~{}}}\mathcal{D}_2  }\left[\frac{\sum_{i=1}^K w_{i}R_i\left(\hat{\mathbf{x}}, \mathbf{G}\right)}{\sum_{i=1}^K w_{i}R_i\left(\hat{\mathbf{x}}_{\text{FP}}, \mathbf{G}\right) }\right], \label{eq:performance}
\end{align}
where $\hat{\mathbf{x}}$ is the output of LUVA with $N$ iterations (i.e. layers) and $\hat{\mathbf{x}}_{\text{FP}}$ is the output of FPLinQ after $100$ iterations for given $\mathbf{G}, \mathbf{w}$. The proposed deep unfolded algorithm LUVA is trained with a fixed number of layers, with $\text{MaxIter} = 1e+6$, training rate $t_r = 1e-3$, refinements $ f_1 = 1e-1, f_2 = 1e-2$ and $n_\text{train} = \lbrace 50, 100, 500 \rbrace$ for training. 

\subsubsection{Initial Value}
As an introduction to the numerical experiments of LUVA, we begin by discussing the choice of the initial value. In the later experiments, the initial value is set to ${\bf x}_0 = 0.01\cdot\bf 1$. It was observed that this choice increases performance, as shown in Table \ref{tab:init_comp}. While an initial points of zero or all ones already produce reasonable results, we found that an initial point where every user starts with $1\%$ of its total power performs best. This may be attributed to the location on the rate region. It is likely that LUVA performs better with ${\bf x}_0 = 0.01\cdot\bf 1$ because we are close to the Pareto-Boundary, but not directly on it, as would be the case with ${\bf x}_0 = \bf 1$. 
\begin{table}[h]
    \centering
    \begin{tabular}{c|c|c|c}
        $n_\text{train}$& ${\bf x}^{(0)}_1 = \bf 0$ & $ {\bf x}^{(0)}_2= 0.01\cdot\bf 1$ & ${\bf x}^{(0)}_3 = \bf 1$ \\ \hline
        $500$& $99.23\%$ & $\bf 100.65\%$ & $98.91\%$
    \end{tabular}
    \caption{Comparisson of the performance of trained LUVA with $N = 10$ for different initial values, tested on 1000 unseen network layouts with $K=10$ user.}
    \label{tab:init_comp}
\end{table}
\subsubsection{Training}
\begin{figure}
    \centering
    \includegraphics[width=0.45\textwidth]{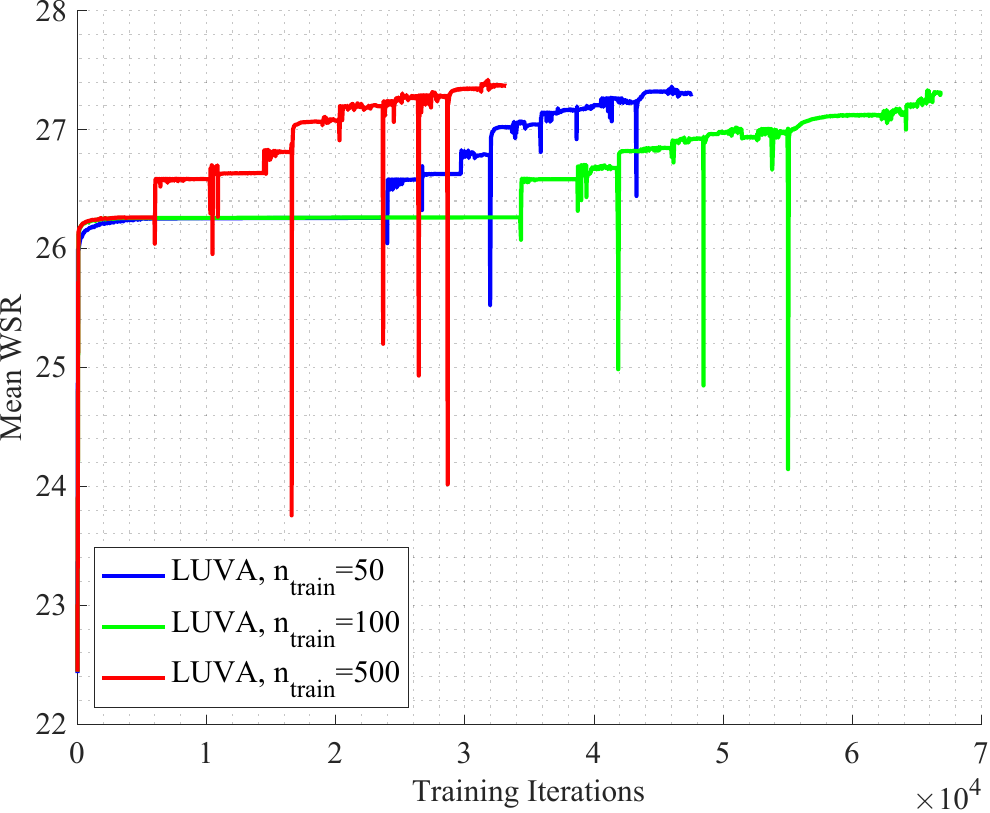}
    \caption{Training history for LUVA with $N=10$ for $n_\text{train} = \lbrace 50, 100, 500\rbrace$ of the (fixed) testing dataset, with $500$ samples, for $K = 10$. The observed jumps come from going one layer to another, thus we see this \textit{staircase} behaviour.}
    \label{fig:train_hist}
\end{figure}
In Figure \ref{fig:train_hist} the training history of the validation set for $K = 10$ and $N=10$ is presented for a different number of training samples per iteration. The three curves show the mean weighted sum-rate of the proposed algorithm on the unseen validation data-set with $500$ samples and random uniform weights. In each case jumps and peaks are observed in the training history, these usually occurs when jumping from one layer/iteration to the next one. Moreover one can already observe that the algorithm trained with more training samples per training-iteration can achieve a higher mean weighted sum-rate. This is to be expected as more training data allows the trainable parameters to adapt to a larger set of D2D-networks. Furthermore it can be observed that in the latter iterations the jumps get smaller as we reach an (local) optimum. 
\subsubsection{Performance}
\begin{figure}[h!]
    \centering
    \includegraphics[width=1.\linewidth]{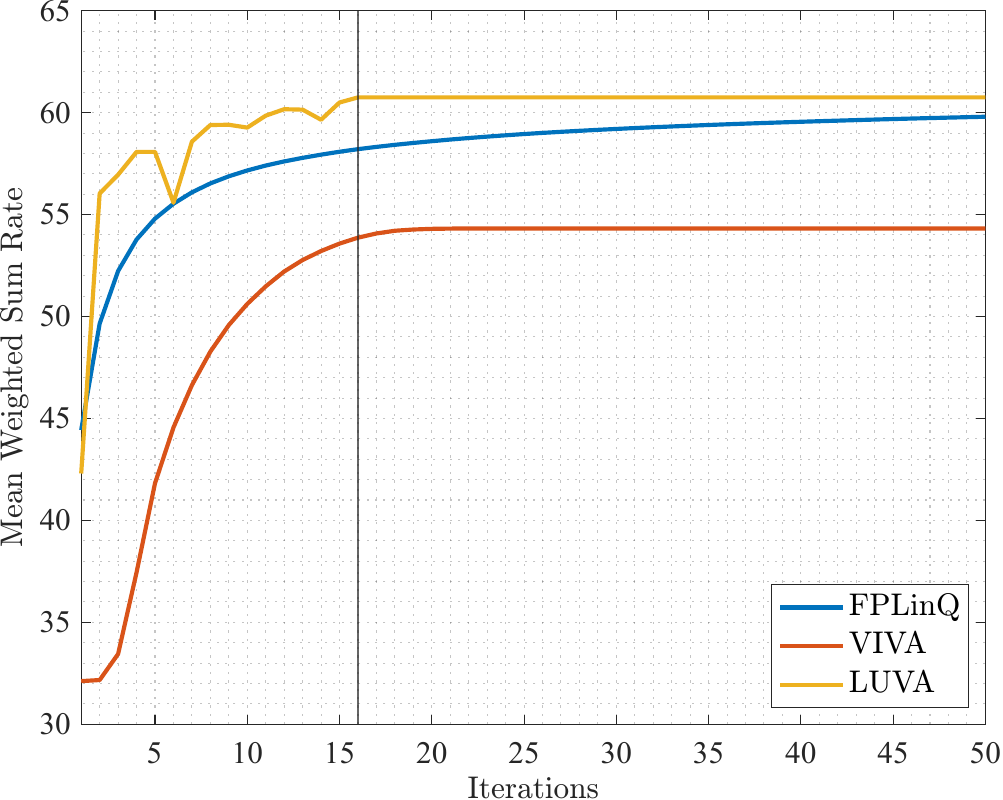}
    \caption{The mean weighted-sum-rate versus iterations of the benchmark FPLinQ and both the trained and untrained versions of the proposed Algorithm \ref{LearnedPD} is evaluated across 1000 unseen D2D networks with $K=50$ users and random uniform weights. LUVA is trained with $n_\text{train}=500$, using random uniform weights and $N=16$ iterations, as indicated by the vertical black line in the figure. After the 16th iteration, no further iterations were performed, and to provide a visual representation, the final value was repeated. It can be observed that VIVA becomes stuck at a sub-optimal point, while the trained version successfully escapes these optima. Remarkably, LUVA achieves a mean weighted-sum-rate higher than FPLinQ, already by the 16th iteration.}
    \label{fig:sum_vs_trained_conv}
\end{figure}
We will continue with investigating the overall performance of LUVA. For this LUVA is trained once with fixed weights equal to one and for random uniform weights $w_i\text{ \raisebox{-0.9ex}{\~{}} }\mathcal{U}_{[0,1]}$ for every generated training and validation D2D network-layout with $K = 50$ users. In Figure \ref{fig:sum_vs_trained_conv} the faster convergence by employing deep unfolding can be experimentally shown. Moreover one can observe that the optimized parameters are able to help the algorithm to escape local optima and thus the trained algorithm is able to achieve the same mean weighted-sum-rate FPLinQ achieves but already after $16$ Iterations. In Table \ref{tab:LVA_vanilla} the results for $K=50$ users are presented for a different number of samples used in each training-iteration, tested on 1000 unseen layouts. The table shows that LUVA, trained with random uniform weights and $n_\text{train} = 500$, is able to achieve the same performance of FPLinQ for less iterations and thus less complexity. 

Moreover, the trained algorithm is able to perform as well as FPLinQ when the weights are equal to one. However, using only weights equal to one during training decreases the overall performance and is not well generalizable to problems where the weights are drawn random uniformly. This outcome is expected since the cone $\mathcal{K}_{\bf w}$ becomes more complex when the weights are randomly uniformly distributed and this variability was not encountered during training. 

\begin{figure}
    \centering
    \includegraphics[width=0.45\textwidth]{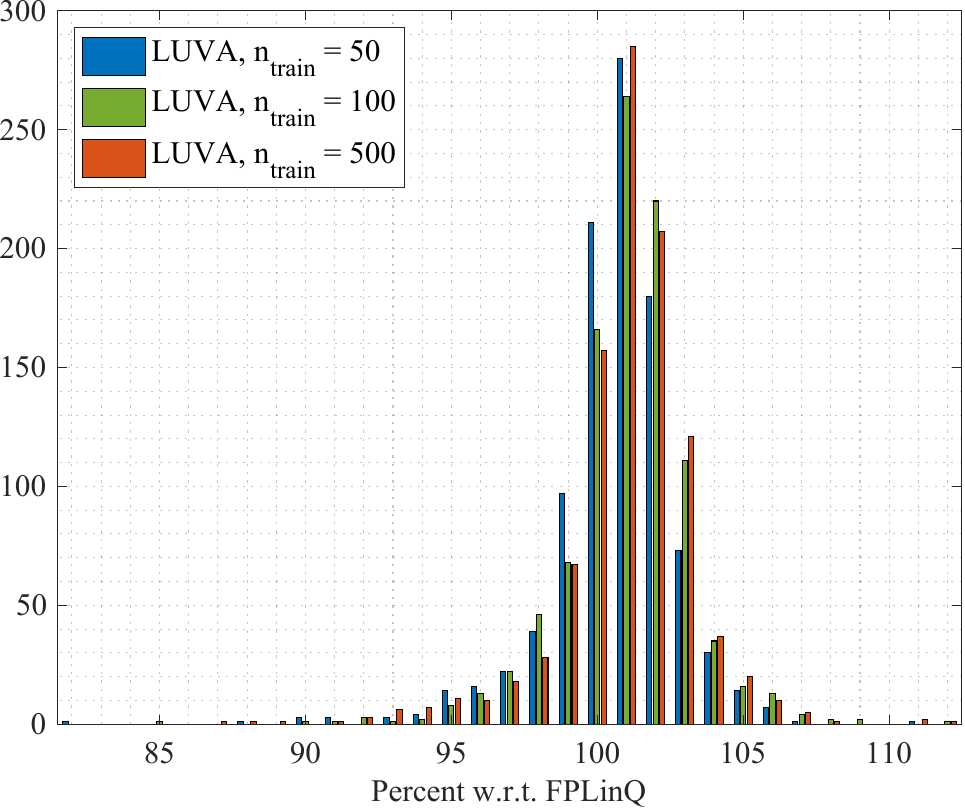}
    \caption{Histogram of LUVA with $N=16$ and $n_\text{train} \in\lbrace50, 100, 500\rbrace$ w.r.t. FPLinQ for 1000 unseen test layouts with $K=50$ links. Trained and tested with random uniform weights.}
    \label{fig:histo}
\end{figure}

Additionally to the latter experiments the histogram of the experiment is shown in Figure \ref{fig:histo}. We see in all cases a concentration around $100\%$ percent. One can observe, a larger number of training samples per training-iteration is increasing the number of cases where LUVA is outperforming FPLinQ. In total we have $797$ cases greater ore equal to $100\%$ for LVA$(N = 16)$ with $n_\text{train} = 50$ while with $n_\text{train} = 100$ we have $834$. With $n_\text{train} = 500$ we have $846$ cases above or equal $100\%$. This means in the setting of $N=16$ layers and $\lbrace 100, 500\rbrace$ training samples per iteration we have only $\approx 15\%$ of unseen D2D-setting where LUVA performs worse than $100\%$ of the achieved instantaneous weighted sum-rate achieved by FPLinQ, while having a lower complexity. 

FPLinQ can be viewed as a fixed-point algorithm based on the first-order optimality condition of \eqref{eq:weightedSum}, making it entirely parameter-free, see \cite{shen_fractional_2018}. In contrast, the proposed unfolded algorithm LUVA is based on gradient descent and does rely on parameters. However, this parameter dependency is turned into an advantage: through deep unfolding, the parameters are optimized to accelerate convergence. As a result, the unfolded algorithm requires fewer iterations than FPLinQ, thereby lowering computational complexity.

\begin{table}[h!]
    \centering
    \begin{tabular}{c|c|c|c|c}
    & \multicolumn{2}{c|}{trained on $w_i\text{ \raisebox{-0.9ex}{\~{}} }\mathcal{U}_{[0,1]}$} & \multicolumn{2}{c}{trained on $\mathbf{w}=\mathbf{1}$} \\\hline
    $n_\text{train}$ & $w_i\text{ \raisebox{-0.9ex}{\~{}} }\mathcal{U}_{[0,1]}$& $\mathbf{w}=\mathbf{1}$ & $w_i\text{ \raisebox{-0.9ex}{\~{}} }\mathcal{U}_{[0,1]}$& $\mathbf{w}=\mathbf{1}$ \\ \hline
    $50$ & $100.1\%$ & $99.22\%$ & $91.37 \%$& $ 98.79\%$  \\ \hline
    $100$ & $100.5\%$ & $98.97$ & $92.08\%$& $98.83\%$   \\ \hline
    $500$ & ${\bf 100.53\%}$& ${\bf 100.1\%}$ & $95.80\%$ & $99.70\%$ 
\end{tabular}

    \caption{Performance \eqref{eq:performance} of LUVA with $N=16$ and trained with random uniform weights for each network (left) and trained on weights equal to one (right).}
    \label{tab:LVA_vanilla}
\end{table}
\subsubsection{Generalizability}
\begin{table}
    \centering
    \begin{tabular}{c|c|c|c}x
    Size & $K$ & $w_i\text{ \raisebox{-0.9ex}{\~{}} }\mathcal{U}_{[0,1]} $ & $\mathbf{w}=\mathbf{1}$  \\ \hline
    $250$ m $\times$ $250$ m & $13$ & $ 100.3 \%$ & $100.3\%$\\ \hline
    $500$ m $\times$ $500$ m & $50$ & ${\bf 100.5\%}$ & $\bf 100.1\%$ \\ \hline
    $750$ m $\times$ $750$ m & $113$ & $100.1$ &  $99.36 \%$\\ \hline
    $1$ km $\times$ $1$ km & $200$ & $ 98.76\%$&  $98.2\%$\end{tabular}

    \caption{Performance \eqref{eq:performance} of LUVA with $N=16$ and, trained on $K=50$ on an area of $500$ m $\times$ $500$ m and with random uniform weights, tested on settings with same link density. .}
    \label{tab:same_density}
\end{table}
\begin{figure*}[htp]
    \centering 
    \begin{subfigure}{0.45\linewidth}
        \centering
        \includegraphics[width=1.\textwidth]{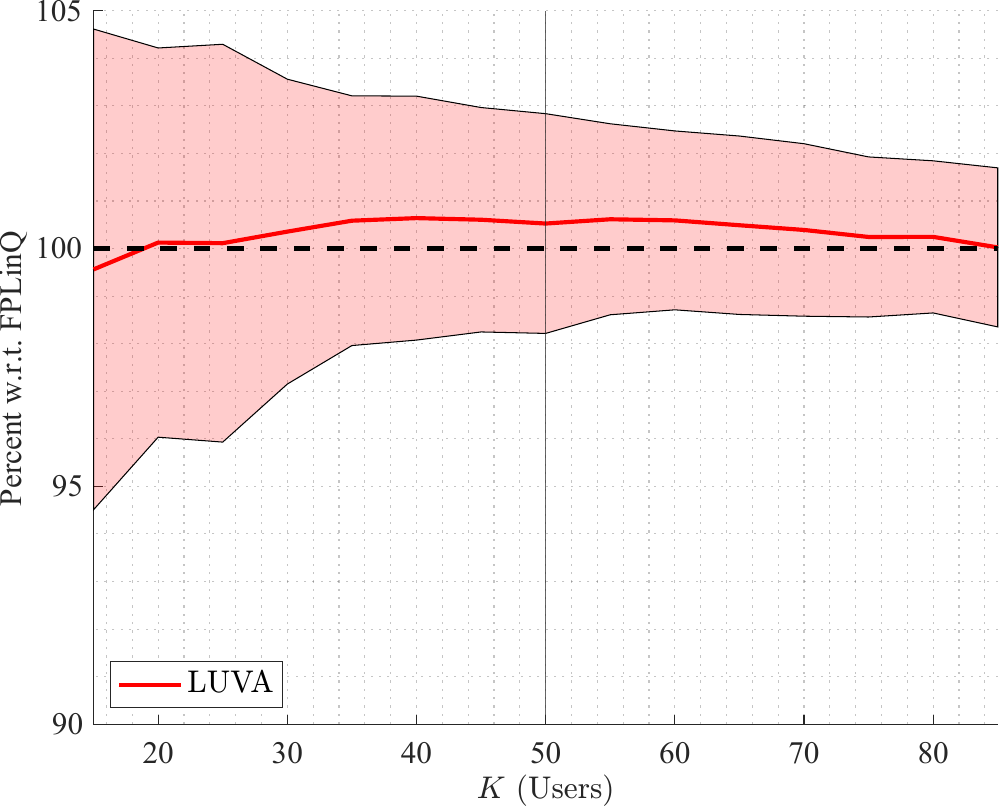}
         \caption{Performance \eqref{eq:performance} of proposed LUVA with $N=16$ trained on settings with different link densities in terms of mean and standard deviation w.r.t. the mean FPLinQ sum rate. The vertical line indicates the training setting.}
        \label{fig:work_interval}
    \end{subfigure}
    \hskip\baselineskip
    \begin{subfigure}{0.45\linewidth}
        \centering
        \includegraphics[width=1.\textwidth]{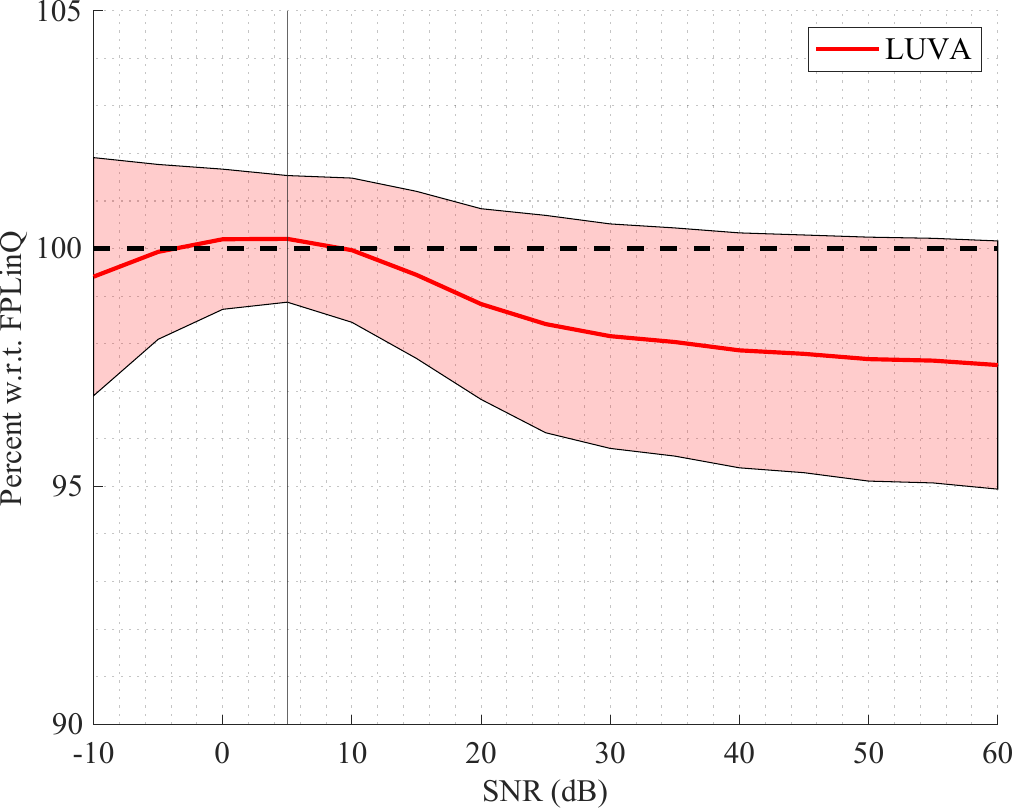}
        \caption{Performance \eqref{eq:performance} of proposed LUVA with $N=16$ trained on settings with different SNR in terms of mean and standard deviation w.r.t. the mean FPLinQ sum rate. The vertical line indicates the training setting.}
        \label{fig:SNR_test}
    \end{subfigure}
    \vskip\baselineskip
    \begin{subfigure}{0.45\linewidth}
        \centering
        \includegraphics[width=1.\textwidth]{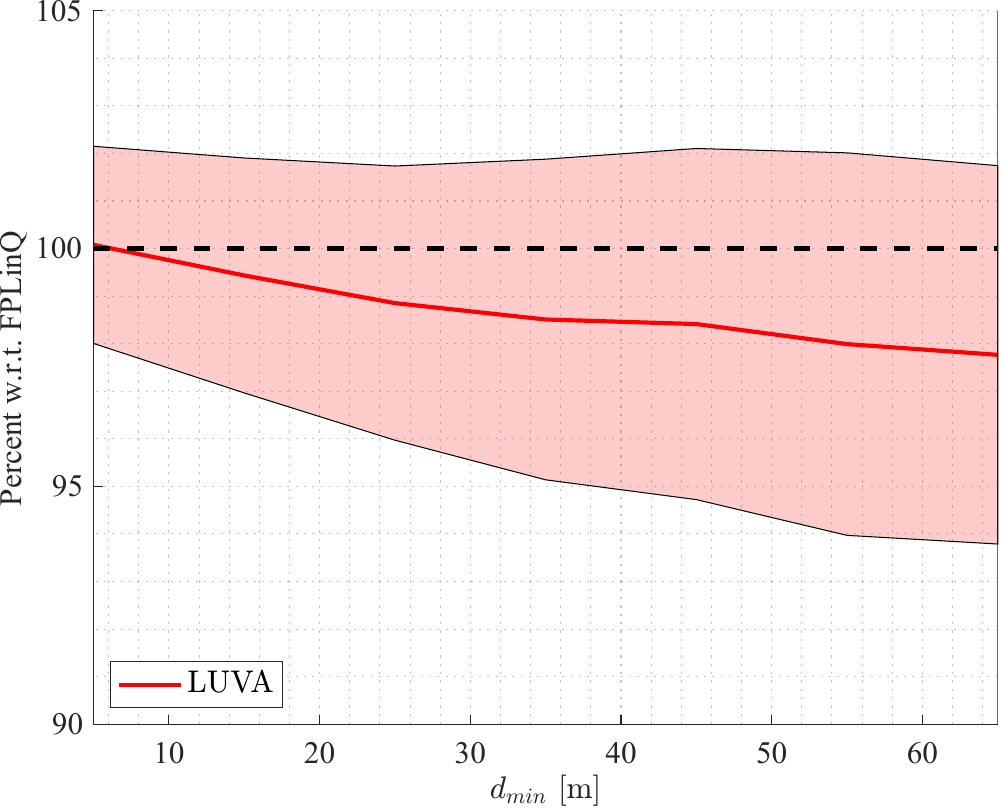}
        \caption{Performance \eqref{eq:performance} of proposed LUVA with $N=16$ on settings with different $d_\text{min}$.}
        \label{fig:length_test_d_min}
    \end{subfigure}
    \hskip\baselineskip
    \begin{subfigure}{0.45\linewidth}
        \centering
        \includegraphics[width=1.\textwidth]{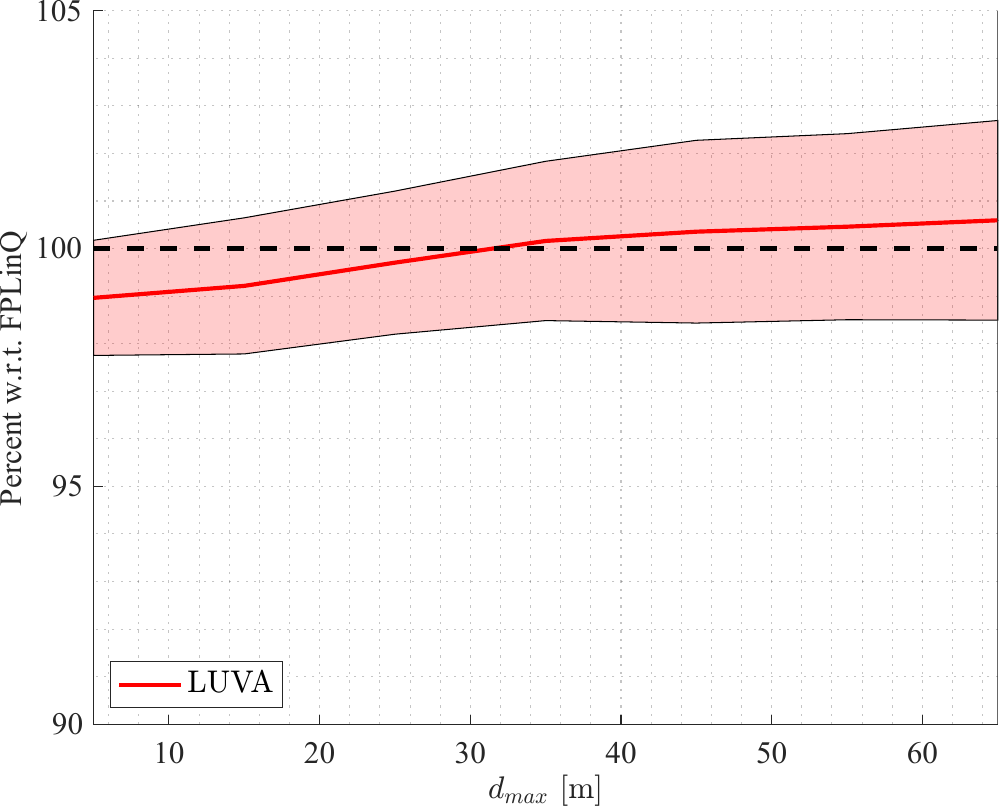}
        \caption{Performance \eqref{eq:performance} of proposed LUVA with $N=16$ on settings with different $d_\text{max}$.}
        \label{fig:length_test_d_max}
    \end{subfigure}
    \caption{Experiments w.r.t. the generalizability of the proposed unfolded algorithm LUVA with $N=16$, for $K=50$ and random uniform weights: a) by adding or subtracting users to the D2D-networks, b) by changing the SNR level and c), d) by changing the $d_\text{min}, d_\text{max}$ while the other one is fixed to default value. The vertical lines in a), b) indicate the training setting.}
    \label{fig:gen_ex}
\end{figure*}
One important aspect of data-driven approaches is their generalizability to unseen distributions. For this LUVA trained on random uniform weights for $K=50$ and $n_\text{train} = 500$ is considered. We investigate the behavior of the trained algorithm on scenarios with consistent link density, scenarios with a fixed area and varying number of links $K$ (thus changing link density), and different SNR levels. As previously mentioned, the proposed method is independent of $K$. Since only scalars are being trained, the trained model can thus be easily applied to settings with a larger or smaller number of users.

In Table \ref{tab:same_density} LUVA with $N=16$, trained with $K=50$, random uniform weights and $n_\text{train} = 500$, is tested on areas with different sizes and fixed link density. It can be observed that the network trained with random uniform weights is able to generalize very well, as the performance is almost in all cases $\geq 100\%$. On the other hand one can observe a slow drop in performance, when the number of links increases significantly, which could indicate that more iterations are needed at this point, since the problem gets more complex as $K$ increases.

In Figure \ref{fig:gen_ex} the experiments w.r.t. the generalizability of the proposed LUVA are presented. In each subfigure the red line indicates the performance \eqref{eq:performance} w.r.t. FPLinQ as before, additionally, the red-shaded area shows the standard deviation of the discrete expected value given in \eqref{eq:performance}.

Figure \ref{fig:work_interval} shows the proposed LUVA performance tested on D2D-networks with a different number of links, while the size of the area is fixed to $500\text{ m}\times500\text{ m}$, trained for $K=50$ and random uniform weights for each D2D-network. Here we can observe that the proposed LUVA almost always reaches a performance of  $\approx 100$\%. 

Overall, the latter results show, that it is sufficient to save a set of trained network parameters and deploy them only, if the number of additional or leaving users exceeds a certain number.

In Figure \ref{fig:SNR_test}, we investigate the effect of different SNR levels on performance. For this experiment, we changed the transmit power of the underlying D2D network and retrained the algorithm. The SNR shown in Figure \ref{fig:SNR_test} is obtained by placing one link at the maximum distance without interference, the algorithm is retrained for SNR$=5$dB. It can be observed that the proposed LUVA is sensitive to changes in SNR levels, as performance declines when the SNR is increased or decreased above $\pm5$ dB. On the other hand we still have a performance of over $95\%$.

Finally, in Figure \ref{fig:length_test_d_min} and \ref{fig:length_test_d_max} we investigate the proposed LUVA on D2D network layouts with different $d_\text{min}, d_\text{max}$ than to the training topology. Specifically, the minimum distance $d_\text{min}$ of each receiver to their respective transmitter is increased, while $d_\text{max}$ is fixed to the default value of $65$ m. Analogously this is done for the experiments w.r.t. $d_\text{max}$, while here $d_\text{min}$ is fixed as $2$ m, as in the training data and $d_\text{max}$ is decreased. In both cases one can observe that the performance decreases if $d_\text{min}\to d_\text{max}$ and vice versa. This is likely because the distance between each link becomes equal, leading to system matrices that differ significantly from those generated during the training procedure. Especially if $d_\text{min}$ increases we see a larger drop in performance, since the interference will increase, if the distance between transmitter and receiver pairs increases.

In the latter experiments, it could be shown that the proposed LUVA is as good as the benchmark FPLinQ, with the same complexity per iteration, but fewer iterations itself, thus achieving the same performance as FPLinQ for a lower overall complexity. Furthermore, it was observed that once trained, the unfolded network is able to generalize to settings with fewer or more users without losing too much performance compared to FPLinQ. Thus, one could use already trained parameters when the number of leaving and joining D2D links does not exceed a certain threshold. Moreover the performance of the unfolded algorithms decreases in settings where the interference changes stronger to the training scenario, this was observed in Figures \ref{fig:SNR_test}, \ref{fig:length_test_d_min} and \ref{fig:length_test_d_max}. In these cases the algorithm has to be re-trained, respectively another set of trained parameters has to be deployed.
\subsection{Network Utility Maximization}
For now LUVA was only tested on weights equal to one or random uniform weights. We could show experimentally that the unfolded algorithm is able to generalize well to different weights and scenarios. In practice weights are used to impose priority on certain links, for example, a link with very low gain could be inactive for a long time or not be activate at all. Thus, the link should get a weight assigned, s.t. it is higher prioritized in some next time slot. This can be formulated as a \textit{Network Utility Maximization Problem} (NUM problem) and then iteratively approximated over discrete time slots. More specifically, the objective of the NUM problem is able to control the long-term throughput rates of each user. By the general method of virtual queues and Lyapunov
Drift Plus Penalty (DPP) one is able to update the weights for each time slot, s.t. the throughput rates of each link converge to the solution of the NUM problem. In this section, it is experimentally shown that the proposed LUVA, already trained on random uniform weights, used as the solver for the WSRM in each discrete time slot is competitive to FPLinQ and can be used as a building block in this farines scheduler. The NUM-problem is usually defined as follows. 
The long-term average throughput for each user $i$ is defined as
\begin{align*}
    \bar{R}_i(\mathbf{G}) = \lim_{T\to\infty} \frac{1}{T} \sum_{t = 1}^T R_i(\mathbf{x}(t), \mathbf{G})
\end{align*}
where $R_i(\mathbf{x}(t), \mathbf{G})$ is the rate of user $i$ at discrete time slot $t$, w.r.t. the power control variable $\mathbf{x}(t)$, for a given D2D network with system matrix $\mathbf{G}$. The NUM-Problem is then stated as
\begin{align}
    \max_{\bar{\mathbf{R}} \in\mathcal{R}(\mathbf{G}) } U(\bar{\mathbf{R}}(\mathbf{G}) )\label{eq:NUM}
\end{align}
where $\mathcal{R}(\bf G)$ is the convex hull of all achievable rates of the given D2D network and $U(\cdot)$ is a concave and (component-wise) non decreasing function. Note that \eqref{eq:NUM} represents a convex optimization problem; however, the difficulty in efficiently solving it arises because $\mathcal{R}(\mathbf{G})$ becomes computationally infeasible to calculate for a large number of users. In the context of link selection, that is ${\bf x}\in\lbrace0,1\rbrace^K$, the number of possible activation combinations grows exponentially, resulting in a linear combination of $2^K - 1$ achievable rates. In general $\mathcal{R}(\mathbf{G})$ is more complex.

Therefore, the weights are updated according to the method of virtual queues and Lyapunov Drift Plus Penalty (DPP), \cite{neely2010stochastic}. Specifically the weights are updated as
\begin{align*}
    Q_{i}(t+1, \mathbf{G}) = \left[Q_{i}(t, \mathbf{G}) - R_i(\mathbf{x}(t), \mathbf{G})\right]_+ + A_i(t, \mathbf{G}),
\end{align*}
for all $i = 1,\dots,K$, where $\mathbf{A}(t, \mathbf{G}) = \mathbf{a}$ and $\mathbf{a}$ is the solution of the following auxiliary convex problem 
\begin{align*}
    \max_{\mathbf{a}} VU(\mathbf{a}) - \sum_{i=1}^KQ_i(t, \mathbf{G})a_i\\
    \text{s.t. }\mathbf{a}\in [0, A_\text{max} ]^K
\end{align*}
and $\mathbf{R}(\mathbf{x}(t), \mathbf{G})$ is the solution of the (instantaneous) weighted-sum-rate problem, with weights $Q_i(t, {\bf G})$, i.e.
\begin{align}
    \max_\mathbf{x} \sum_i^KQ_i(t, \mathbf{G})R_i(\mathbf{x}(t), \mathbf{G}).\label{eq:NUM_rate}
\end{align}
It is known, \cite{neely2010stochastic}, that the convergence of this mechanism satisifies
\begin{align*}
    \liminf_{T\to\infty} U\left( \frac{1}{T}\sum_{t=1}^T \mathbf{R}(\mathbf{x}(t), \mathbf{G})\right) &\geq U(\bar{\mathbf{R}}^*(\mathbf{G})) - \frac{\kappa}{V} \\ 
    \limsup_{T\to\infty} \frac{1}{T} \sum_{t = 1}^T Q_i(t, {\bf G}) &= \mathcal{O}(V)\, \forall\,i
\end{align*}
where $\kappa$ is a system dependent constant and $\bar{\mathbf{R}}^*(\mathbf{G})$ is the solution of \eqref{eq:NUM}, for any $Q_i(0, {\bf G})$. In particular, as V increases, the DPP algorithm approaches the optimal solution of \eqref{eq:NUM} within a gap that decreases a $\mathcal{O}(1/V)$, while the convergence time increases as $\mathcal{O}(V)$. 

Note that this mechanism is very sensitive w.r.t. the choice of $V$, as the $Q_i$'s fluctuate too much for a small $V$, but if $V$ is chosen too large, they do not converge to a steady state in a reasonable time. The choice of $V$ is done heuristically in the next experiments. Moreover, $A_\text{max}$ is chosen s.t. the hypercube $[0, A_\text{max} ]^K$ contains $\mathcal{R}(\mathbf{G})$ and is thus given as the maximum element of the Utopian point \eqref{eq:Utopian}, i.e. $A_\text{max} = \max_i u_i$.

In the following, the performance of this scheme is evaluated for networks with $K=10$ user and parameters given in Table \ref{tab:system_params}, where FPLinQ or the proposed LUVA are used as building blocks to solve the weighted-sum-rate for given weights $Q_i(t)$.

The discussed DPP algorithm is run separately for the proposed LUVA as well as for FPLinQ to have a fair comparison, since the weights $\mathbf{Q}(t+1)$ depend on the decision of the optimizer. LUVA is pre-trained offline on random uniform weights and for $N=10$ iterations and for $K=10$ user and a training set size of $n_\text{train} = 500$. 

In the case of LUVA, the weights are normalized w.r.t. the maximum-norm, to have a more consistent input, this is the only pre-processing step applied on the calculated weights. 

We consider two objective functions, one aiming to achieve proportional fairness and one to achieve max-min fairness, \cite{mo2000fair}. In both cases the discussed scheme is carried out for $1000$ time slots and the throughput rates for each user are computed by taking the last $500$ time slots. This is done to ensure convergence of the DPP algorithm to a steady point. 
 \begin{figure*}[htp]
    \centering 
    \begin{subfigure}{0.45\linewidth}
        \centering
        \includegraphics[width=1.\textwidth]{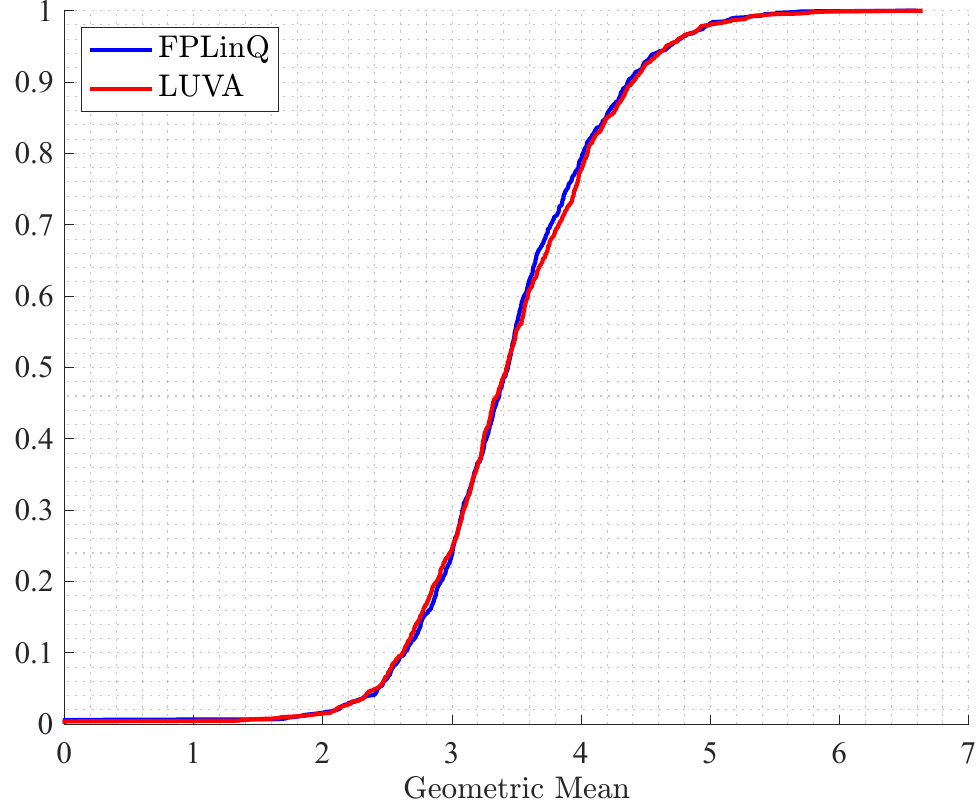}
        \caption{CDF of the GM of achieved throughput rates by FPLinQ and LUVA for \eqref{eq:NUM}, with  $U(\bar{\bf R})=\sum_i^K \log(\bar{R}_i)$ with  $V = 10$.}
        \label{fig:NUM_PF}
    \end{subfigure}
    \hskip\baselineskip
    \begin{subfigure}{0.45\linewidth}
        \centering
        \includegraphics[width=1.\textwidth]{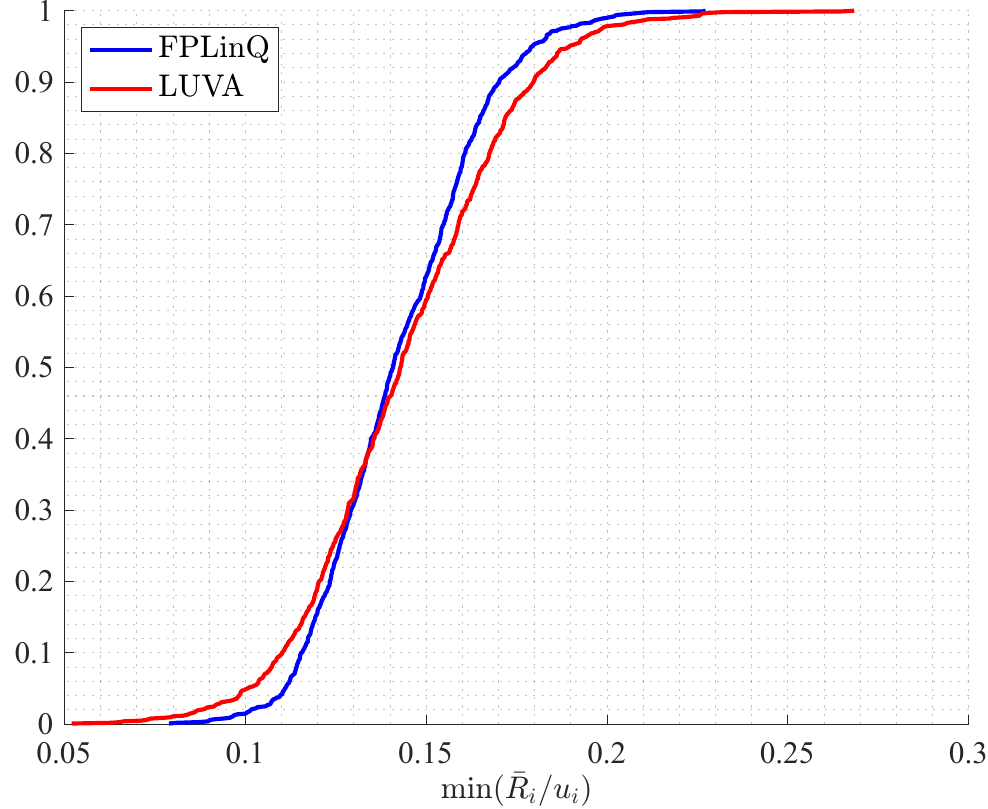}
        \caption{CDF of FPLinQ, LUVA and the upper bound for \eqref{eq:NUM} with $U(\bar{\bf R})=\min_{i = \lbrace1,\dots,K\rbrace}\bar{R}_i/ u_i$, with $V = 40$.}
        \label{fig:NUM_W-HF}
    \end{subfigure}
    \caption{CDF of the NUM-Problem \eqref{eq:NUM} objective, which is given as a) $U(\bar{\bf R})=\sum_i^K\log(\bar{R}_i)$ and b) $U(\bar{\bf R})=\min_{i = \lbrace1,\dots,K\rbrace}\bar{R}_i/u_i$. Experiments where done for $1000$ random generated D2D networks with $K = 10$ user. The throughput rates for each user are taken over the last $T=500$ time slots. Moreover, LUVA is trained for $N=10$ layers / iterations and $K=10$ users on random uniform weights and with $n_\text{train} = 500$. While in a) the performance of the proposed LUVA is almost exact to the benchmark, in b) there is always a lower value obtained, w.r.t. FPLinQ, by considering max-min fairness.}
    \label{fig:NUM_1}
\end{figure*}
\subsubsection{Proportional Fairness} 
To enforce \textit{proportional fairness} along all links the utility function is chosen as
\begin{align*}
    U\left(\bar{\bf R}\right) = \sum_{i=1}^K \log(\bar{R}_i).
\end{align*}
It is shown in \cite{mo2000fair} that the solution, $\bar{\bf R}^*$, to \eqref{eq:NUM} with the latter objective is equivalent with the point fulfilling the aggregate proportional change criterion, defined as
\begin{align}
    \sum_{i=1}^K \frac{\bar{R}_i - \bar{R}_i^*}{\bar{R}_i^*} \leq 0,\, \forall\bar{\bf R}\in\mathcal{R}(\mathbf{G}).
\end{align}

The results can be observed in Figure \ref{fig:NUM_PF} in form of the Cumulative Distribution Function (CDF) of the geometric mean,
\begin{align*}
    \text{GM}(\bar{\bf R}) = \left(\prod_{i = 1}^K \bar{R}_i\right)^{1/K},
\end{align*}
obtained over $1000$ D2D network realizations (unseen in the training phase). Here the results obtained by FPLinQ (blue) and the proposed method (red) are shown. It can be seen that the CDF of FPLinQ and LUVA almost coincide, indicating the effectiveness of the proposed LUVA. 
\subsubsection{Weighted Max-Min Fairness}
By the utility function given as
\begin{align}\label{eq:W-HF}
    U\left(\bar{\bf R}\right) = \min_{i={\lbrace1,\dots,K\rbrace}}\frac{\bar{ R}_i}{{u}_i}.
\end{align}
one enforces that the ratio of throughput of rates to their maximum peak rate are almost the same across all links. Figure \ref{fig:NUM_W-HF} shows the results w.r.t. \textit{weighted max-min fairness} again in the form of the CDF of the objective in \eqref{eq:W-HF} for each randomly generated D2D topology. Here, it can be observed again, that the performance of the proposed LUVA is almost coinciding with the one of FPLinQ. 

Overall it can be observed that the proposed LUVA is able to achieve a similar performance as FPLinQ, with a lower complexity, as a building block in a fairness scheduler. Moreover, LUVA is trained offline with random uniform weights and run here with weights derived by the virtual queue DPP algorithm, which are generally not random and uniform, which again shows it's good generalizability to different weights. 

\section{Conclusion}
In this work, we proposed an alternative problem formulation to solve the weighted-sum-rate maximization problem for dense D2D networks. A primal-dual algorithm is then developed to address the new problem formulation, its parameters can be optimized by interpreting each iteration as the layer of a neural network, this approach is known as deep unfolding. In numerical experiments it could be shown that the proposed unfolded algorithm is able to reach a comparable performance w.r.t. the benchmark with fewer iterations and has thus a lower complexity.
The trained algorithm could also show strong generalizability w.r.t. changing the number of users, different SNR regimes and arbitrary weights. Finally, the proposed unfolded algorithm could shown to be an efficient building block in a dynamic fairness scheduling scheme to maximize a concave network utility function of the long-term user throughput rates. 



{\footnotesize
  \bibliographystyle{IEEEtran}
  \bibliography{bib}

\begin{thebibliography}{10}
\providecommand{\url}[1]{#1}
\csname url@samestyle\endcsname
\providecommand{\newblock}{\relax}
\providecommand{\bibinfo}[2]{#2}
\providecommand{\BIBentrySTDinterwordspacing}{\spaceskip=0pt\relax}
\providecommand{\BIBentryALTinterwordstretchfactor}{4}
\providecommand{\BIBentryALTinterwordspacing}{\spaceskip=\fontdimen2\font plus
\BIBentryALTinterwordstretchfactor\fontdimen3\font minus
  \fontdimen4\font\relax}
\providecommand{\BIBforeignlanguage}[2]{{%
\expandafter\ifx\csname l@#1\endcsname\relax
\typeout{** WARNING: IEEEtran.bst: No hyphenation pattern has been}%
\typeout{** loaded for the language `#1'. Using the pattern for}%
\typeout{** the default language instead.}%
\else
\language=\csname l@#1\endcsname
\fi
#2}}
\providecommand{\BIBdecl}{\relax}
\BIBdecl

\bibitem{mo2000fair}
J.~Mo and J.~Walrand, ``Fair end-to-end window-based congestion control,''
  \emph{IEEE/ACM Transactions on networking}, vol.~8, no.~5, pp. 556--567,
  2000.

\bibitem{neely2010stochastic}
M.~J. Neely, ``Stochastic network optimization with application to
  communication and queueing systems,'' \emph{Synthesis Lectures on
  Communication Networks}, vol.~3, no.~1, pp. 1--211, 2010.

\bibitem{wu_flashlinq_nodate}
X.~Wu, S.~Tavildar, S.~Shakkottai, T.~Richardson, J.~Li, R.~Laroia, and
  A.~Jovicic, ``Flashlinq: A synchronous distributed scheduler for peer-to-peer
  ad hoc networks,'' \emph{IEEE/ACM Transactions on Networking}, vol.~21,
  no.~4, pp. 1215--1228, 2013.

\bibitem{naderializadeh_itlinq_2014}
N.~Naderializadeh and A.~S. Avestimehr, ``Itlinq: A new approach for spectrum
  sharing in device-to-device communication systems,'' \emph{IEEE journal on
  selected areas in communications}, vol.~32, no.~6, pp. 1139--1151, 2014.

\bibitem{geng_optimality_2015}
\BIBentryALTinterwordspacing
C.~Geng, N.~Naderializadeh, A.~S. Avestimehr, and S.~A. Jafar,
  ``\BIBforeignlanguage{en}{On the {Optimality} of {Treating} {Interference} as
  {Noise}},'' \emph{\BIBforeignlanguage{en}{IEEE Transactions on Information
  Theory}}, vol.~61, no.~4, pp. 1753--1767, Apr. 2015. [Online]. Available:
  \url{http://ieeexplore.ieee.org/document/7051266/}
\BIBentrySTDinterwordspacing

\bibitem{yi_optimality_2015}
X.~Yi and G.~Caire, ``Optimality of treating interference as noise: A
  combinatorial perspective,'' \emph{IEEE Transactions on Information Theory},
  vol.~62, no.~8, pp. 4654--4673, 2016.

\bibitem{chiang2007power}
M.~Chiang, C.~W. Tan, D.~P. Palomar, D.~O'neill, and D.~Julian, ``Power control
  by geometric programming,'' \emph{IEEE transactions on wireless
  communications}, vol.~6, no.~7, pp. 2640--2651, 2007.

\bibitem{shen_fplinq_2017}
\BIBentryALTinterwordspacing
K.~Shen and W.~Yu, ``\BIBforeignlanguage{en}{{FPLinQ}: {A} cooperative spectrum
  sharing strategy for device-to-device communications},'' in
  \emph{\BIBforeignlanguage{en}{2017 {IEEE} {International} {Symposium} on
  {Information} {Theory} ({ISIT})}}.\hskip 1em plus 0.5em minus 0.4em\relax
  Aachen, Germany: IEEE, Jun. 2017, pp. 2323--2327. [Online]. Available:
  \url{http://ieeexplore.ieee.org/document/8006944/}
\BIBentrySTDinterwordspacing

\bibitem{shen_fractional_2018}
\BIBentryALTinterwordspacing
------, ``\BIBforeignlanguage{en}{Fractional {Programming} for {Communication}
  {Systems}—{Part} {I}: {Power} {Control} and {Beamforming}},''
  \emph{\BIBforeignlanguage{en}{IEEE Transactions on Signal Processing}},
  vol.~66, no.~10, pp. 2616--2630, May 2018. [Online]. Available:
  \url{http://ieeexplore.ieee.org/document/8314727/}
\BIBentrySTDinterwordspacing

\bibitem{shen_fractional_2018-1}
\BIBentryALTinterwordspacing
------, ``\BIBforeignlanguage{en}{Fractional {Programming} for {Communication}
  {Systems}—{Part} {II}: {Uplink} {Scheduling} via {Matching}},''
  \emph{\BIBforeignlanguage{en}{IEEE Transactions on Signal Processing}},
  vol.~66, no.~10, pp. 2631--2644, May 2018. [Online]. Available:
  \url{https://ieeexplore.ieee.org/document/8310563/}
\BIBentrySTDinterwordspacing

\bibitem{cui_spatial_2019}
\BIBentryALTinterwordspacing
W.~Cui, K.~Shen, and W.~Yu, ``\BIBforeignlanguage{en}{Spatial {Deep} {Learning}
  for {Wireless} {Scheduling}},'' \emph{\BIBforeignlanguage{en}{IEEE Journal on
  Selected Areas in Communications}}, vol.~37, no.~6, pp. 1248--1261, Jun.
  2019. [Online]. Available: \url{http://arxiv.org/abs/1808.01486}
\BIBentrySTDinterwordspacing

\bibitem{lee_graph_2020}
\BIBentryALTinterwordspacing
M.~Lee, G.~Yu, and G.~Y. Li, ``\BIBforeignlanguage{en}{Graph {Embedding} based
  {Wireless} {Link} {Scheduling} with {Few} {Training} {Samples}},''
  \emph{\BIBforeignlanguage{en}{arXiv:1906.02871 [cs, eess]}}, Nov. 2020,
  arXiv: 1906.02871. [Online]. Available: \url{http://arxiv.org/abs/1906.02871}
\BIBentrySTDinterwordspacing

\bibitem{zhao2022link}
Z.~Zhao, G.~Verma, C.~Rao, A.~Swami, and S.~Segarra, ``Link scheduling using
  graph neural networks,'' \emph{IEEE Transactions on Wireless Communications},
  2022.

\bibitem{shelim_geometric_2022}
R.~Shelim and A.~S. Ibrahim, ``Geometric {Machine} {Learning} {Over}
  {Riemannian} {Manifolds} for {Wireless} {Link} {Scheduling},'' \emph{IEEE
  Access}, vol.~10, pp. 22\,854--22\,864, 2022, conference Name: IEEE Access.

\bibitem{levie_radiounet_2021}
\BIBentryALTinterwordspacing
R.~Levie, C.~Yapar, G.~Kutyniok, and G.~Caire,
  ``\BIBforeignlanguage{en}{{RadioUNet}: {Fast} {Radio} {Map} {Estimation}
  {With} {Convolutional} {Neural} {Networks}},''
  \emph{\BIBforeignlanguage{en}{IEEE Transactions on Wireless Communications}},
  vol.~20, no.~6, pp. 4001--4015, Jun. 2021. [Online]. Available:
  \url{https://ieeexplore.ieee.org/document/9354041/}
\BIBentrySTDinterwordspacing

\bibitem{sun_learning_2018}
\BIBentryALTinterwordspacing
H.~Sun, X.~Chen, Q.~Shi, M.~Hong, X.~Fu, and N.~D. Sidiropoulos,
  ``\BIBforeignlanguage{en}{Learning to {Optimize}: {Training} {Deep} {Neural}
  {Networks} for {Interference} {Management}},''
  \emph{\BIBforeignlanguage{en}{IEEE Transactions on Signal Processing}},
  vol.~66, no.~20, pp. 5438--5453, Oct. 2018. [Online]. Available:
  \url{https://ieeexplore.ieee.org/document/8444648/}
\BIBentrySTDinterwordspacing

\bibitem{lee2018deep}
W.~Lee, M.~Kim, and D.-H. Cho, ``Deep learning based transmit power control in
  underlaid device-to-device communication,'' \emph{IEEE Systems Journal},
  vol.~13, no.~3, pp. 2551--2554, 2018.

\bibitem{chen2018convlista}
\BIBentryALTinterwordspacing
X.~Chen, J.~Liu, Z.~Wang, and W.~Yin, ``Theoretical linear convergence of
  unfolded ista and its practical weights and thresholds,'' in \emph{Advances
  in Neural Information Processing Systems}, S.~Bengio, H.~Wallach,
  H.~Larochelle, K.~Grauman, N.~Cesa-Bianchi, and R.~Garnett, Eds.,
  vol.~31.\hskip 1em plus 0.5em minus 0.4em\relax Curran Associates, Inc.,
  2018. [Online]. Available:
  \url{https://proceedings.neurips.cc/paper/2018/file/cf8c9be2a4508a24ae92c9d3d379131d-Paper.pdf}
\BIBentrySTDinterwordspacing

\bibitem{li_graph-based_2022}
B.~Li, G.~Verma, and S.~Segarra, ``Graph-based algorithm unfolding for
  energy-aware power allocation in wireless networks,'' \emph{IEEE Transactions
  on Wireless Communications}, vol.~22, no.~2, pp. 1359--1373, 2022.

\bibitem{shen2019graph}
Y.~Shen, Y.~Shi, J.~Zhang, and K.~B. Letaief, ``A graph neural network approach
  for scalable wireless power control,'' in \emph{2019 IEEE Globecom Workshops
  (GC Wkshps)}.\hskip 1em plus 0.5em minus 0.4em\relax IEEE, 2019, pp. 1--6.

\bibitem{pascoletti1984scalarizing}
A.~Pascoletti and P.~Serafini, ``Scalarizing vector optimization problems,''
  \emph{Journal of Optimization Theory and Applications}, vol.~42, pp.
  499--524, 1984.

\bibitem{el2011network}
A.~El~Gamal and Y.-H. Kim, \emph{Network information theory}.\hskip 1em plus
  0.5em minus 0.4em\relax Cambridge university press, 2011.

\bibitem{eichfelder_adaptive_2008}
\BIBentryALTinterwordspacing
G.~Eichfelder, \emph{\BIBforeignlanguage{en}{Adaptive Scalarization Methods in
  Multiobjective Optimization}}, ser. Vector {Optimization}, J.~Jahn, Ed.\hskip
  1em plus 0.5em minus 0.4em\relax Berlin, Heidelberg: Springer Berlin
  Heidelberg, 2008. [Online]. Available:
  \url{http://link.springer.com/10.1007/978-3-540-79159-1}
\BIBentrySTDinterwordspacing

\bibitem{zadeh_weighted_sum}
L.~Zadeh, ``Optimality and non-scalar-valued performance criteria,'' \emph{IEEE
  Transactions on Automatic Control}, vol.~8, no.~1, pp. 59--60, 1963.

\bibitem{haimes1971bicriterion}
Y.~Haimes, ``On a bicriterion formulation of the problems of integrated system
  identification and system optimization,'' \emph{IEEE transactions on systems,
  man, and cybernetics}, no.~3, pp. 296--297, 1971.

\bibitem{khaledian2016restricting}
K.~Khaledian, ``Restricting the parameter set of the pascoletti-serafini
  scalarization,'' \emph{Bulletin of the Iranian Mathematical Society},
  vol.~42, no. 7 (Special Issue), pp. 89--112, 2016.

\bibitem{liu2019alista}
J.~Liu and X.~Chen, ``Alista: Analytic weights are as good as learned weights
  in lista,'' in \emph{International Conference on Learning Representations
  (ICLR)}, 2019.

\bibitem{hauffen2022algorithm}
J.~C. Hauffen, P.~Jung, and N.~M{\"u}cke, ``Algorithm unfolding for
  block-sparse and mmv problems with reduced training overhead,''
  \emph{Frontiers in Applied Mathematics and Statistics}, vol.~9, p. 1205959,
  2023.

\bibitem{7934066}
M.~Borgerding, P.~Schniter, and S.~Rangan, ``Amp-inspired deep networks for
  sparse linear inverse problems,'' \emph{IEEE Transactions on Signal
  Processing}, vol.~65, no.~16, pp. 4293--4308, 2017.

\bibitem{osman_AMP_2021}
O.~Musa, P.~Jung, and G.~Caire, ``Plug-and-play learned gaussian-mixture
  approximate message passing,'' in \emph{ICASSP 2021-2021 IEEE International
  Conference on Acoustics, Speech and Signal Processing (ICASSP)}.\hskip 1em
  plus 0.5em minus 0.4em\relax IEEE, 2021, pp. 4855--4859.

\bibitem{8550778}
Y.~Yang, J.~Sun, H.~Li, and Z.~Xu, ``Admm-csnet: A deep learning approach for
  image compressive sensing,'' \emph{IEEE Transactions on Pattern Analysis and
  Machine Intelligence}, vol.~42, no.~3, pp. 521--538, 2020.

\bibitem{miriya2022deep}
U.~S. Miriya~Thanthrige, P.~Jung, and A.~Sezgin, ``Deep unfolding of
  iteratively reweighted admm for wireless rf sensing,'' \emph{Sensors},
  vol.~22, no.~8, p. 3065, 2022.

\bibitem{kingma2014adam}
D.~P. Kingma and J.~Ba, ``Adam: A method for stochastic optimization,''
  \emph{arXiv preprint arXiv:1412.6980}, 2014.

\bibitem{chen_hyperparameter_2021}
X.~Chen, J.~Liu, Z.~Wang, and W.~Yin, ``Hyperparameter tuning is all you need
  for lista,'' \emph{Advances in Neural Information Processing Systems},
  vol.~34, pp. 11\,678--11\,689, 2021.

\bibitem{georgiadis2006resource}
L.~Georgiadis, M.~J. Neely, L.~Tassiulas \emph{et~al.}, ``Resource allocation
  and cross-layer control in wireless networks,'' \emph{Foundations and
  Trends{\textregistered} in Networking}, vol.~1, no.~1, pp. 1--144, 2006.

\bibitem{neely2012stability}
M.~J. Neely, ``Stability and probability 1 convergence for queueing networks
  via lyapunov optimization,'' \emph{Journal of Applied Mathematics}, vol.
  2012, no.~1, p. 831909, 2012.

\end{thebibliography}
}


 




\vfill

\end{document}